\documentclass[aps,prd,groupedaddress,amsmath,amssymb,amsfonts,nofootinbib,preprint,longbibliography]{revtex4-1}
\usepackage{amssymb,amsfonts}
\usepackage{mathrsfs}
\usepackage{tikz}
\usetikzlibrary{external}
\usepackage{amsthm}
\usepackage{xcolor}
\usepackage{graphicx}
\usepackage[utf8x]{inputenc}
\usepackage{float}
\usepackage{amsmath}
\usepackage{mathtools}
\usepackage{bbold}
\usepackage{accents}
\usepackage{cancel}
\usepackage{esint}
\DeclareMathAlphabet{\mathpzc}{OT1}{pzc}{m}{it}
\usetikzlibrary{patterns}
\usepackage{url}
\usepackage{cleveref}
\pdfoutput=1
\usepackage{epsfig}
\usepackage{graphics}
\begin{document}

\title{Gravitational Effects of  Null Particles}
\author{Kris Mackewicz}
\affiliation{University of Chicago}
\author{Craig Hogan}
\affiliation{University of Chicago}
\date{\today}
\begin{abstract}
We generalize earlier solutions of gravitational shocks sourced by distributional null stress-energy, and analyze their observable effects. A systematic method is proposed to demonstrate consistency between the gravitational effects of a null perfect fluid, i.e. a photon gas, and a collection of gravitational shock waves from point null particles.
The  anisotropy and time dependence of observable gravitational shifts produced by individual null shocks  on a spherically arranged system of clocks relative to a central observer are derived. Shock solutions are used
to characterize the angular and temporal spectra of gravitational fluctuations  from a photon gas, inherited from the micro structure of the individual null particles.  Angular and temporal frequency spectra and correlation functions of gravitational redshift are computed   as functions of the ratio of particle impact parameter and transverse size to the size of the measuring sphere.
These results are applied to estimate observable large-scale correlations of  
gravitational fluctuations  produced by thermal  states of null particles. A pure-phase component of anisotropy is computed from differential gravitational  time shifts. It is argued that this component may provide clues to the angular and temporal structure of measurable gravitational vacuum fluctuations. 
\end{abstract}
\maketitle

\section{Introduction}
Solutions to the Einstein field equations for null sources of stress-energy have been of considerable interest since the seminal work of  \citep{Aichelburg1971}. 
The most symmetric solutions are elegantly simple: the gravitational effect produced by a null plane  is  characterized as a discontinuous displacement in space-time relationships on  opposite sides of the null planar shock\cite{DRAY1985173}.

    In this paper, we generalize and adapt these solutions to describe situations that are more physically realistic. The first exercise is to show explicitly that  a system that approaches the continuum limit of many null shocks produces the same gravitation as an idealized relativistic  gas. We then use the solutions to characterize the concrete physical effects of individual shocks, via measurements of a spatially extended array of clocks  viewed by an observer on a single world line. This analysis is then applied to estimate classical gravitational fluctuations produced by a physical photon gas, as well as  gravitational fluctuations produced by a quantum  vacuum.  These results include spectra and correlation functions in frequency,  angular wave number, and angular separation, which will be useful in the design and interpretation of experiments that seek to measure quantum-gravitational fluctuations.

Distributional sources of stress-energy  present computational difficulties since the Green's function integral methods are non-convergent. Stress-energy sources of distributional character  present particular challenges since the Einstein equations are non-linear, and products of distributions are not always mathematically well defined. For this reason, it is often convenient to work with the linearized Einstein equations, although there is some ambiguity in what is meant by ``small'' when handling terms proportional to delta functions or their derivatives. There have been several works  characterizing the behavior of solutions involving distributional stress-energy \cite{David_Garfinkle_1999,Geroch87,Tolish:2014a,Israel1991,DRAY1985173,CBarrabes_1989,poisson2002reformulation}. It has been shown that in some instances, it is possible to solve for the full non-linear solutions to the Einstein equations that have distributional curvature.

Here, we explicitly outline a procedure for averaging the spacetime curvature of a collection of gravitational shock waves sourced by null point particles. In one application,  we  show that the acceleration of test bodies in a perfect null fluid can be consistently described as a series of instantaneous velocity kicks in random directions due to the passing of many gravitational shock waves. 
Then, we characterize the angular and temporal spectrum of fluctuations of the gravitational effects of these null fluids, starting with the effects of a single shock wave crossing an array of spherically arranged clocks, similar to the setup outlined in \cite{Mackewicz_2022}.

We begin by outlining in detail the solution to the linearized Einstein equations for a perfect null fluid with equation of state $P=\rho/3$ in the Lorenz gauge. We then compute the curvature of such a system and characterize the motion of geodesics in this spacetime. 
Next, we study in detail the solution to the linearized Einstein equations for a single null point-particle, also in the Lorenz gauge. We extend this solution to more generic (that is, extended, non-pointlike) distributions of null stress-energy, and characterize their effect on timelike geodesics. At this point, we provide a systematic method for averaging the metric perturbations (up to possible gauge transformations) and spacetime curvatures for a collection of randomly oriented null point-particles, and demonstrate that the result is the same effect as one would expect from a perfect null fluid.

We then use the generalized shock  solution to derive   angular and temporal spectra and correlation functions  of gravitational redshift and displacement  measured by a spherically arranged system of clocks due to  single generalized null particles,  as a function of the central impact parameter and transverse size of their energy distribution in the null plane.
These power spectra and correlation functions  describe gravitational fluctuations  from a   gas of  photons. 
We  also isolate the pure-phase component of distortions, which may provide clues to symmetries of gravitational fluctuations from 
vacuum fluctuations.

\section{Gravity of a Photon Gas} 
Let us first investigate the connection between the null point-particle solutions to the Einstein equations and the perfect continuum null fluid (photon gas) solution in the linearized regime.

A convenient model to use for studying continuum sources of stress-energy is the ``perfect fluid'' model. This requires only knowledge of an equation of state $P=P(\rho)$ to characterize the system, along with simplifying assumptions such as isotropy and homogeneity. This type of modeling has been used extensively in cosmological models (e.g. FRW spacetimes), characterized by a scale factor $a(t)$ where
\begin{equation}
    ds^2 = -dt^2 + a(t)^2(dx^2+dy^2+dz^2). \label{FRWmetric}
\end{equation}
Solving the Friedmann equations, one can show that $a(t)\propto t^{2/{3(w+1)}}$, where $w$ defines the equation of state $P=w \rho$\cite{weinberg2008cosmology}. On smaller scales, these models can be used to approximate stellar formation in the linearized/Newtonian regime (e.g. Jeans instability). One can also study the dynamics of the so called ``photon gas'' using this model. This is what is commonly referred to as a ``radiation dominated'' FRW universe.

A photon gas can be modeled as a perfect fluid with equation of state $P=\rho/3$ (working in units where $c=1$) 
\begin{equation}
    T_{ab} = (\rho + P)u_au_b+P g_{ab}. \label{perfectfluidstress}
\end{equation}
In an orthonormal tetrad, this becomes
\begin{equation}
    T_{ab} = \rho t_at_b+\frac{1}{3}\rho \delta_{ab}
\end{equation}
where $\delta_{ab}$ here and henceforth represents the flat 3D spatial metric. The full non-linear Einstein equations give
\begin{align}
    R = -8\pi T = 0 \\
    R_{ab} = 8 \pi T_{ab}
\end{align}
The solution to the non-linear Einstein equations is the FRW metric (eq. \ref{FRWmetric}) of a radiation dominated universe with the scale factor evolving as $a(t)\propto t^{1/2}$, and $\rho(t)\propto a(t)^{-4}\propto t^{-2}$ \cite{weinberg2008cosmology}. 

We will focus on making comparisons in the linearized regime, since one can only treat the delta functions present in the shock waves solutions meaningfully in this limit. Consider the linearized Einstein equations in the Lorenz gauge $\nabla^a\bar{h}_{ab}=0$:
\begin{equation}
    \Box \bar{h}_{ab}=-16\pi T_{ab}=-16\pi\rho\left(t_at_b+\frac{1}{3}\delta_{ab}\right) .\label{linearizedeinstein}
\end{equation}
It is usually customary to work in Newtonian gauge for this type of model, but comparisons between the results of the photon gas and the shock waves can be made more readily when working in the same gauge. If one tries to enforce a time-independent metric perturbation in the Lorenz gauge, then the resulting solution is not isotropic. To see this, first let
\begin{equation}
    \bar{h}_{ab} = f(\vec{x})\left(t_at_b+\frac{1}{3}\delta_{ab}\right).
\end{equation}
To have a self-consistent solution, we need
\begin{align}
    \nabla^2 f = -16\pi \rho \label{laplacef} \\
    \delta_{ab}\nabla^bf = 0 \label{lorenzf}.
\end{align}
However, one can clearly see that if we take the divergence of eq. (\ref{lorenzf}), it gives $\nabla^2f=0$, which is inconsistent with eq. (\ref{laplacef}) and therefore leads to a contradiction. A similar contradiction arises if we try to make $f$ purely a function of time. To remedy this issue, we must have
\begin{equation}
    \bar{h}_{ab} = f(\vec{x})t_at_b+g(t)\delta_{ab}
\end{equation}
such that 
\begin{align}
    \nabla^2 f &= -16\pi \rho \label{laplacef2}\\
    \partial_t^2 g &= \frac{16\pi}{3}\rho \label{laplaceg}
\end{align}
Solving eq. (\ref{laplacef2})-(\ref{laplaceg}), we find
\begin{equation}
    \bar{h}_{ab} = -\frac{8\pi}{3}\rho (r^2t_at_b -t^2\delta_{ab}) \label{gasmetric}
\end{equation}
\begin{equation}
    \bar{h}=\bar{h}_{ab}\eta^{ab}=\frac{8\pi}{3}\rho (r^2+3t^2)
\end{equation}
Recalling that $h_{ab}=\bar{h}_{ab}-\frac{1}{2}\eta_{ab}\bar{h}$, we find 
\begin{equation}
    h_{ab} = -\frac{4\pi}{3}\rho \left((r^2-3t^2)t_at_b + (r^2+t^2)\delta_{ab}\right)
\end{equation}
The linearized Riemann curvature tensor is then
\begin{equation}
    R_{abcd} = -\frac{16\pi}{3} \rho \delta_{[d|[a}\left(2t_{b]|}t_{c]} + \delta_{b]|c]}\right) \label{gascurvature}
\end{equation}
Consistent with the FRW models (which are known to be conformally flat), the Weyl tensor vanishes in the linearized regime as well:
\begin{align}
    C_{abcd}&=R_{abcd} - \frac{1}{2}(R_{ac}g_{bd}-R_{ad}g_{bc}-R_{bc}g_{ad}+R_{bd}g_{ac})-\frac{R}{6}(g_{ad}g_{bc}-g_{ac}g_{bd}) \\
    &=-\frac{16\pi}{3} \rho \delta_{[a[d}\left(2t_{b]}t_{c]} + \delta_{b]c]}\right) +16\pi\rho (t_{[a}t_{[d}+\frac{1}{3}\delta_{[a[d})\eta_{c]b]}\\
    &=0. \label{Weylcurvature}
\end{align}

If one examines the linearized geodesic deviation equation, we find that nearby timelike geodesics accelerate relative to one-another by the relation
\begin{equation}
    \frac{d^2D^a}{dt^2}\approx -R_{tbt}^{\quad a}D^b = -\frac{8\pi}{3}\rho D^a, \label{geodesic deviation}
\end{equation}
where $D^a$ represents the deviation vector between two nearby geodesics that are initially parallel. It is worth pointing out that if $\rho=const$, as would be the case in the strictly linear regime to satisfy conservation of stress energy, then nearby timelike geodesics tend to be focused. However, if we consider non-linear corrections, $\rho$ becomes a decreasing function of time, and nearby timelike geodesics now tend to drift apart rather than focus.

Now that we have characterized the spacetime for the homogeneous and isotropic null fluid, we focus our attention on the gravitational shockwaves sourced by null point particles. 

\section{Gravitational Shockwaves}
Gravitational shockwaves can be characterized by spacetimes with metrics that contain a discontinuity across a null hypersurface and/or have singular curvature components on this null hypersurface. After the seminal papers describing the shocks\cite{Aichelburg1971,DRAY1985173},  several authors  developed standard techniques for joining spacetimes with particular boundary conditions, leading to propagating shockwaves \cite{Israel1991, CBarrabes_1989,poisson2002reformulation}. Such spacetimes have also been studied in the context of gravitational memory\cite{Tolish:2014a}. 

The types of shockwaves we are primarily interested in here are those which have flat geometry to the causal past and present of the shockwave, and the only curvature effects are concentrated in a null plane with infinite spatial extent perpendicular to the direction of propagation. As mentioned in \cite{Tolish:2014a}, there are some mathematical difficulties treating such solutions using retarded Green's function methods, and one should really think of this solution as a limit of a spherical shockwave that was created infinitely far in the past by some kind of emission process\footnote{In fact, there is no L2 or distributional solution for the metric perturbation strictly speaking, but one can write down a non-distributional solution that does give rise to the correct curvature, which is a well defined distribution.}. The important feature of these kinds of waves is that they give rise to an instantaneous relative velocity kick (a form of memory) to nearby test particles.

\subsection{Single Photon Shockwave}

The standard method for determining the linearized metric perturbation for a non-null source is to invert the linearized Einstein equations using the 3D retarded Green's function for the wave operator \citep{Wald:1984}. In other words, we seek solutions to the equation
\begin{equation}
    \bar{h}_{ab}=4\int \frac{T_{ab}(x',t_{ret}(x'))}{|x-x'|}d^3x' \label{invertedwaveeq}
\end{equation}
where $\bar{h}_{ab}$ is the trace-reversed metric perturbation in the Lorenz gauge. However, for null point particles that exist for all time, there is an issue of convergence of the integral since the intersection of the past light cone with the source is infinite. To find the solution, the original authors carefully took the limit as $v \rightarrow 1$ of the boosted Schwarzschild solution \cite{Aichelburg1971}. The stress-energy of a point particle with 4-velocity $u^a$ is given by
\begin{equation}
    T_{ab} = \gamma^{-1} m \delta^{(3)}(\vec{x}-\vec{X}(t))u_au_b \label{pointparticlestress}
\end{equation}
This leads to a metric perturbation in the Lorenz gauge of
\begin{equation}
    h_{ab} = \frac{2m}{ \gamma\alpha|\vec{x}-\vec{X}(t_{\rm ret})|}(2u_au_b+\eta_{ab}) \label{pointparticlemetric}
\end{equation}
where we define $\alpha$ as in \cite{2022ace..book.....W}:
\begin{equation}
    \alpha = 1 - \hat{n}\cdotp \dot{\vec{X}}(t_{\rm ret})
\end{equation}
\begin{equation}
    \hat{n} = \frac{\vec{x}-\vec{X}(t_{\rm ret})}{|\vec{x}-\vec{X}(t_{\rm ret})|}
\end{equation}
Taking $\vec{X}(t)=v t \hat{z}$ and taking the limit $v\rightarrow 1$ for $t\neq z$ gives
\begin{equation}
    ds^2 = - dt^2 + d\vec{x}^2 +\frac{4E}{|t-z|}(dt-dz)^2 \quad (t\neq z). \label{shocklineelement}
\end{equation}
This metric is flat everywhere except the null hypersurface $t=z$, and does not give us a direct method of computing the correct curvature on the null hypersurface. The authors in \citep{Aichelburg1971} propose a clever trick for taking the $v\rightarrow 1$ limit, and the result is\footnote{As pointed out by the original authors, one should be cautious when working with a metric of distributional character since products of metric components are not well-defined and the Einstein equations are non-linear. However, $det(g)$ and $g^{-1}$ can be defined and the linearized Riemann curvature tensor gives an answer consistent with the full non-linear solution for the curvature using the boosted Schwarzschild metric.} 
\begin{equation}
    \bar{h}_{ab}=h_{ab} = -8 E \,{\rm ln}(s) \delta(t-z) k_a k_b \label{shockmetric}
\end{equation}
where $k_a=t_a+z_a=-\nabla_a u$. The associated curvature is given by
\begin{equation}
    R_{abcd} = \frac{16E}{s^2}\delta(u)k_{[a}(s_{b]}s_{[c}-\phi_{b]}\phi_{[c})k_{d]} - 16 \pi E \delta(x)\delta(y)\delta(u)k_{[a}q_{b][c}k_{d]} \label{shockcurvature}
\end{equation}
where $s_a,\phi_a$ represent unit vectors in cylindrical polar coordinates $(t,s,\phi,z)$ and $q_{ab}$ is the 2D spatial projection onto the plane transverse to the direction of propagation of the photon. The first term in eq. (\ref{shockcurvature}) is traceless and is the Weyl curvature, while the second term is the Ricci part of the curvature. The geodesic deviation equation for $s\neq 0$ reads
\begin{equation}
    \frac{d^2D_a}{dt^2}\approx \frac{4E}{s^2}\delta(u)(s_{a}s_{b}-\phi_{a}\phi_{b})D^b
\end{equation}
which gives rise to an instantaneous relative velocity kick.

Instead of taking the speed of light limit of the boosted Schwarzschild solution, one can instead consider an {\it ansatz} where the solution satisfies the homogeneous (source free) wave equation in the $t,z$ coordinates, while satisfying a Poisson type equation in the transverse plane ($x,y$ coordinates). 
Consider the following stress-energy tensor representing a null point particle:
\begin{equation}
    T_{ab} = E \delta(t-z)\delta(x)\delta(y) k_a k_b \label{photonstress}
\end{equation}
Assuming the solution is homogeneous in the $t,z$ coordinates, i.e. $\bar{h}_{ab}= f(s) \delta(t-z)k_ak_b$, one just needs to solve
\begin{equation}
    \nabla_{\perp}^2 f(s) = -8 E \delta(s) 
\end{equation}
This differential equation admits the solution
\begin{equation}
    f(s)=-8E\,{\rm ln}(s)
\end{equation}
consistent with the result in eq. (\ref{shockmetric}). 

\subsection{Generic Null Stress-Energy}

Consider a stress energy tensor which represents a superposition of plane waves moving in the $z$ direction, with a varying profile in the $x,y$ plane. 
\begin{equation}
    T_{ab} = E f(u) g(x,y) k_a k_b \label{nullstress}
\end{equation}
It can be easily verified that a stress-energy tensor of this form still satisfies conservation of stress-energy $\nabla^aT_{ab}=0$. In these coordinates, the wave operator can be expanded as
\begin{align}
    \Box &=-\partial_t^2+\partial_z^2+\partial_x^2+\partial_y^2  \\
    &= -4\partial_u\partial_v + \frac{1}{s}\partial_s(s\partial_s) + \frac{1}{s^2}\partial_{\phi}^2 \\
    &= -4\partial_u\partial_v +\nabla_{\perp}^2 \label{waveop}
\end{align}
Combining eq. (\ref{waveop}) and (\ref{linearizedeinstein}) we get
\begin{equation}
    \nabla^2_{\perp} \bar{h}_{ab} = -16 \pi E f(u) g(x,y) k_a k_b \label{transversewaveeq}
\end{equation}
The 2D Green's function for the Poisson equation is given by
\begin{equation}
    G_{\perp}(x,x';y,y') = \frac{1}{2\pi}{\rm ln}((x-x')^2+(y-y')^2)^{1/2} \label{2DGreensfunction}
\end{equation}
\begin{equation}
    \nabla_{\perp}^2 G(x,x';y,y') = \delta(x-x')\delta(y-y')
\end{equation}
We can then get a solution for the trace-reversed metric perturbation by integrating the source against the 2D Green's function.
\begin{equation}
    \bar{h}_{ab} = -8 \int T_{ab}(x',y') {\rm ln}(\sqrt{(x-x')^2+(y-y')^2})dx'dy'
\end{equation}
We will assume cylindrical symmetry so that $g(x,y)=g(s)$, where $s=\sqrt{x^2+y^2}$. Then we can easily invert eq. (\ref{transversewaveeq}) by directly integration of the differential equation. 
\begin{equation}
    \bar{h}_{ab} = f(u) h(s) k_ak_b
\end{equation}
\begin{equation}
    \frac{1}{s}\partial_s(s\partial_s h) = -16 \pi E g(s)
\end{equation}
Depending on the distribution of energy in the transverse plane, we can extract a shockwave solution whose curvature varies in the transverse plane. 

\subsubsection*{i. Uniform Disk}
Consider a uniform density to radius $s=a$, such that $g(s)=(1/\pi a^2)\theta(a-s)$. Then
\begin{equation}
    h(s) = -4 \pi \frac{E}{\pi a^2} s^2 \theta(a-s) - (8 E{\rm ln}(s)+C)\theta(s-a)
\end{equation}
where $C$ is chosen to ensure continuity of the metric at $s=a$, but it is purely gauge and does not contribute to the curvature. For $s<a$, the curvature is 
\begin{equation}
    R_{abcd} = -16\pi\frac{E}{\pi a^2}f(u) k_{[a}q_{b][c}k_{d]} \label{diskshockcurvature}
\end{equation}
For $s>a$, the curvature is the same as the first term in eq. (\ref{shockcurvature}). The important thing to notice is that the profile of the curvature in the $t,z$ plane is the same as that of the stress energy, similar to eq. (\ref{shockcurvature}). Eq. (\ref{diskshockcurvature}) also takes an analogous form to the second term in eq. (\ref{shockcurvature}), where the cross section is a uniform disk in the former case and point like in the latter. 

We will be primarily interested in the case where $f(u)$ is constant over some region, and consider an average over propagation directions. There will be three types of tensor products we need to average over the sphere in eq. (\ref{diskshockcurvature}).
\begin{align}
    t_at_d q_{ac}=t_at_d(\delta_{ac}-z_az_c) \\
    t_a z_d q_{ac} = t_az_d(\delta_{ac}-z_az_c) \\
    z_az_d q_{ac}=z_az_d(\delta_{ac}-z_az_c)
\end{align}
The tensor products $t_at_d,\delta_{ac}$ are unaffected by averaging over the sphere. Any tensor product of an odd number of $z_a$ averages to $0$ over the sphere, while $z_az_c$ averages to $\delta_{ac}/3$. Finally, the term with four $z_a$ will vanish after applying the appropriate index antisymmetry of the Riemann tensor. Summing over co-planar shocks in the $t,z$ plane and averaging over directions gives
\begin{equation}
    R_{abcd} = -\frac{16\pi}{3}\rho \delta_{[d|[a}(2t_{b]|}t_{c]}-\delta_{b]|c]})
\end{equation}
consistent with eq. (\ref{gascurvature}). A more detailed derivation of the result will be shown in section IV. 

\subsubsection*{ii. Gaussian Photon}
Another model of interest is a Gaussian distribution of photon energy in the transverse plane, since in reality particles are not truly point-like. This model will be particularly useful when analyzing the angular spectrum of velocity kicks experienced by clocks on the sphere, since the point-like photon shockwave exhibits singular behavior at the poles ($\theta=0,\pi$). Consider a photon of transverse extent $a$ modeled by a Gaussian distribution. The function $g(x,y)$ must satisfy
\begin{equation}
    g(x,y) = \frac{1}{2\pi a^2}e^{-(x^2+y^2)/2a^2}
\end{equation}
Then we have
\begin{equation}
     \frac{1}{s}\partial_s(s\partial_s h) = -\frac{8 E}{a^2}e^{-s^2/2a^2}
\end{equation}
The solution to this differential equation with the boundary condition $h(0)=0$ is
\begin{equation}
    h(s) = 4 E\left(Ei\left(-\frac{s^2}{2a^2}\right)-{\rm ln}\left(\frac{s^2}{2a^2}\right)-\gamma\right)
\end{equation}
Where $Ei(x)$ is the exponential integral function
\begin{equation}
    Ei(x) = - \int_{-x}^{\infty} \frac{1}{t}e^{-t}dt \label{expintegral}
\end{equation}
and $\gamma$ is the Euler gamma constant (this term is purely gauge, but conveniently sets the boundary condition $h(0)=0$). 

For small argument ($s<<a$) this expression approximates to
\begin{equation}
    h(s)= -\frac{2E}{a^2}s^2
\end{equation}
which is what one would expect for a uniform ``plane wave". For large argument ($s>>a$) we get an asymptotic form of
\begin{equation}
    h(s) \approx - 8 E {\rm ln}(s) +C 
\end{equation}
which is consistent with the point-like particle solution. 

The linearized curvature for a Gaussian shockwave is 
\begin{equation}
    R_{abcd} = 16 E \delta(u)k_{[a}\left(-\frac{1}{a^2}e^{-s^2/2a^2}s_{b]}s_{[c}+\frac{1}{s^2}\left(1-e^{-s^2/2a^2}\right)(s_{b]}s_{[c}-\phi_{b]}\phi_{[c})\right)k_{d]}
\end{equation}
The large distance behavior of the curvature falls off like $1/s^2$ in the transverse direction and is trace free, while the short distance behavior is approximately constant in the transverse plane and has non-vanishing trace. 

\section{Continuum Limit}
Generalized to arbitrary direction, the stress energy of a single point null particle can be written as
\begin{equation}
    T_{ab} = E \delta^{(3)}(\vec{x}-\hat{n}t-\vec{x}_0)n_a n_b
\end{equation}
where $n_a=(1,\hat{n})$, such that $\vec{x}(t=0)=\vec{x}_0$. To reach the continuum limit, we must sum over all points $\vec{x}_0$ and all directions $\hat{n}$. First, consider the 1D case. We want to construct a 1D density function $\rho(x)$ from a sum of delta function sources. 
\begin{equation}
    \rho(x)=\sum_{x_0} \rho(x_0)\delta(x-x_0) \rightarrow \int dx_0 \rho(x_0)\delta(x-x_0)
\end{equation}
This can easily be extended to the $3D$ case
\begin{equation}
   \rho(\vec{x})= \sum_{\vec{x}_0} \rho(\vec{x}_0)\delta^{(3)}(\vec{x}-\vec{x}_0) \rightarrow \int  d^3x_0 \rho(\vec{x}_0)\delta^{(3)}(\vec{x}-\vec{x}_0)
\end{equation}
Summing the individual point-like particle stress-energies over all space with a uniform density $\rho(\vec{x})=1/V$ we get 
\begin{equation}
    \sum_{\vec{x}_0}\rho(\vec{x}_0)T_{ab} \rightarrow \frac{E}{V}n_an_b\int d^3x_0 \delta^{(3)}(\vec{x}-\hat{n}t-\vec{x}_0)=\frac{E}{V} n_an_b 
\end{equation}
Summing over directions $\hat{n}$ amounts to integrating over the sphere (i.e. solid angle). Again, we assume uniform density over solid angle such that $\rho(\theta,\phi)=1/4\pi$.
\begin{equation}
    \frac{1}{4\pi}\int d\Omega \hat{n}_{i}\hat{n}_j = \frac{1}{3} \delta_{ij}
\end{equation}
\begin{equation}
    \sum_{\vec{x}_0}\sum_{\hat{n}} T_{ab} =\frac{E}{V}(t_at_b+\frac{1}{3}\delta_{ab}) 
\end{equation}
As expected, the stress-energy of many point-like particles  averages to that of a continuum fluid. Note that the operation of summing point-like stress-energies preserves conservation of stress-energy, and preserves the trace free $(T^a_a=0)$ condition for a null stress-energy.

\begin{figure}[t]
    \centering
    \includegraphics[width=.6\linewidth]{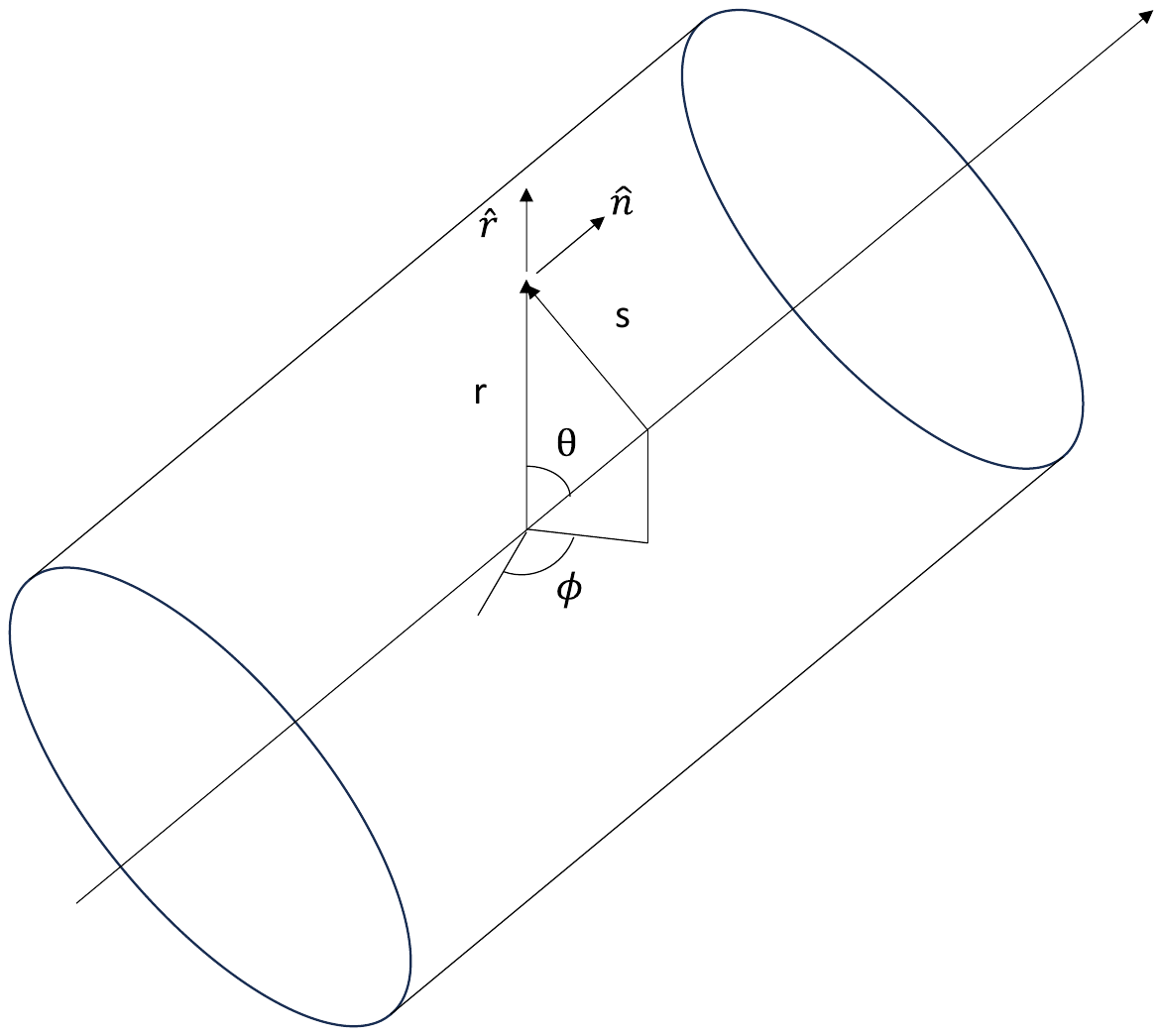}
    \caption{Geometry used for summing shock waves. Each concentric disk forming the cylinder can be though of as a successive shock wave from a bundle of point particles, and we can average over propagation directions to get the metric perturbation and curvature of the photon gas within the ball defined by their intersections.}
    \label{fig:tubegeometry}
\end{figure}

We shall now demonstrate that the same type of summing gives a consistent result for the spacetime metric (up to a gauge transformation) and the Riemann curvature tensor. Summing over points with the same null vector using a uniform spatial density $\rho=1/V$ gives  
\begin{equation}
    \sum_{\vec{x}_0} \bar{h}^{ab} = -4 \pi \frac{E}{V} s^2 k_a k_b \label{tubemetricperturbation}
\end{equation}
We note that eq. (\ref{tubemetricperturbation}) still satisfies the Lorenz gauge condition. Summing over angles using the geometry in fig. \ref{fig:tubegeometry}, we get
\begin{equation}
    \sum_{\vec{x}_0}\sum_{\hat{n}} \hat{h}_{ab} = - 4\pi\frac{E}{V}r^2\left(\frac{2}{3}t_at_b+\frac{4}{15}\delta_{ab}-\frac{2}{15}r_ar_b\right) \label{contmetricpert}
\end{equation}
where $r=s \,{\rm sin}\theta$ is the usual spherical radial coordinate. Using the fact that $\nabla_br_a=(\delta_{ab}-r_ar_b)/r$, one can directly verify that the Lorenz gauge condition $\nabla^a\bar{h}_{ab}=0$ is still satisfied, and we still have that $\bar{h}=h=0$. 

It is immediately apparent that eq. (\ref{contmetricpert}) is not equal to eq. (\ref{gasmetric}). However, one can show that they are equivalent up to a gauge transformation. In general, two metric perturbations give rise to the same curvature if they differ by a gauge transformation of the following form:
\begin{equation}
    h_{ab} \rightarrow h_{ab}+\nabla_a\xi_b+\nabla_b\xi_a
\end{equation}
We want to make a gauge transformation that preserves the Lorenz gauge condition. The necessary condition is that
\begin{equation}
    \Box \xi_a = 0
\end{equation}
In order to recover the metric perturbation from eq. (\ref{gasmetric}), the necessary gauge transformation is
\begin{equation}
    \xi_a = -\frac{2\pi}{3}\frac{E}{V}\left((r^2t+t^3)t_a+(rt^2+\frac{1}{5}r^3)r_a\right)
\end{equation}
We have therefore demonstrated summing over initial photon positions and averaging over propagation directions produces a metric perturbation consistent with that of a photon gas in the Lorenz gauge up to an additional gauge transformation. 

To better understand the significance of this correspondence, let us examine the geodesic deviation equation again:
\begin{equation}
    \frac{d^2D_a}{dt^2} = -R_{bcda}t^bt^dD^c.
\end{equation}
In the point-like photon model, the physically observable effect is an instantaneous relative velocity kick between nearby timelike geodesics given by
\begin{equation}
    \Delta \frac{dD_a}{dt} = \frac{4E}{s^2}(s_as_b-\phi_a\phi_b)D_{(0)}^b - 4\pi E\delta(x)\delta(y)(s_as_b+\phi_a\phi_b)D_{(0)}^b \label{photonrelativekick}
\end{equation}
Recall that $u^a\nabla_aD_b=D^a\nabla_au_b$. Let us consider a perturbative series in powers of $E$ for the four-velocity $u^a$ and the deviation vector $D^a$. In our analysis we will consider a congruence which is initially parallel and consisting of stationary worldlines, so that the deviation vector is initially purely spatial: 
\begin{align}
    u^a&=t^a+u^a_{(1)}\\
    D^a&=D^a_{(0)}+D^a_{(1)}+...
\end{align}
One can easily verify that $u^aD_a=0$ along the entire worldline in consideration. To leading order in $E$, this gives
\begin{equation}
    t^aD_a^{(1)} = -D^a_{(0)}u_a^{(1)} \label{orthogonality}
\end{equation}
Define $v^a=u^b\nabla_b D^a$. By virtue of the fact that $u^b\nabla_b(D^au_a)=0$, one can show that at leading order in $E$ we get
\begin{equation}
    t_av^a_{(1)}=0
\end{equation}
The overall normalization condition $u^au_a=-1$ gives
\begin{equation}
    t_au^a_{(1)}=0 \label{normalization}
\end{equation}
Using eq. (\ref{photonrelativekick})-(\ref{normalization}) we find that a timelike observer picks up a velocity kick after crossing the shock wave given by
\begin{equation}
    \Delta u^a_{(1)} = -\frac{4E}{s}s^a \label{photonkick}
\end{equation}

Let us now consider a superposition of several randomly oriented kicks. If we consider a fixed point of observation and rotate the source position in the transverse plane around the observation point by $\pi/2$, then $\hat{s}\rightarrow -\hat{\phi}$, $\hat{\phi}\rightarrow \hat{s}$. Therefore the first term in eq. (\ref{photonrelativekick}) picks up an overall sign change. If we consider superpositions of co-aligned photon trajectories that are cylindrically symmetrical, all contributions to the relative velocity kick from the first term in eq. (\ref{photonrelativekick}) will cancel out. Summing over the second term for multiple co-aligned trajectories gives the relative velocity kick for a uniform cylindrical disk.
\begin{equation}
    \Delta v_a = -\frac{4\pi E}{A}q_{ab}D^b_{(0)}
\end{equation}
If we smear this kick out over a small interval of time given by $\ell$, then the effective relative acceleration needed to produce this kick is given by
\begin{equation}
    \frac{dv_a}{dt} = -\frac{4\pi E}{V}q_{ab}D^b_{(0)}
\end{equation}
where $V=A\ell$ is the volume of the thin disk over which the energy is spread. Finally, If we average this kick over propagation directions of the photons we get
\begin{equation}
    \frac{dv_a}{dt} = -\frac{8\pi E}{3V} \delta_{ab}D^b{(0)} \label{averagerelativekick}
\end{equation}

In the photon gas model, the physically observable effect is a uniform, isotropic relative acceleration.
\begin{equation}
    \frac{d^2D_a}{dt^2} = - \frac{8\pi}{3}\rho D_a. \label{gasdeviation}
\end{equation}
One can easily check that eq. (\ref{gasdeviation}) is consistent with a coordinate acceleration given by
\begin{equation}
    \frac{du^a}{dt} = -\frac{8\pi}{3}\rho r r^a \label{gasaccel}
\end{equation}
Furthermore, eq. (\ref{gasdeviation}) is equivalent to the averaged effect of a collection of homogeneously and isotropically distributed shock waves due to null point particles given by eq. (\ref{averagerelativekick}). One can similarly show that eq. (\ref{photonkick}) can be summed using Green's function methods in the 2D plane and averaged over directions on the sphere to produce eq. (\ref{gasaccel}) using the fact that 
\begin{equation}
    \nabla_a\left(\frac{s^a}{s}\right) = \frac{\delta(s)}{s}
\end{equation}

\section{Non-Linearity}
In this section, we address the issue of non-linearity of the full solutions to the Einstein field equations. For the photon gas, both the linearized solution and the non-linear (FRW) solutions are well understood. The issue then remains to understand how to treat the non-linearity when dealing with the impulsive gravitational shock waves produces by null point particles. Strictly speaking, we cannot use the null point particle solutions to iterate order by order as the metric and curvature of these solutions are both distributional, and the product of two distributions evaluated at the same spacetime point is ill-defined. However, we propose a series of arguments that suggest that in the continuum limit, understanding the non-linearities of the vacuum portion of the shock waves is not necessary due to the high degree of symmetry of the FRW cosmological solutions. 

\subsection{Expansion}
In this subsection, we demonstrate how all expansion effects, and consequently the dilution of matter, have no dependence on the extended gravitational shockwave emanating from the null point particles (these shockwaves source purely Weyl curvature away from the photon). 
The Raychaudhuri equations gives us the rate of change of the expansion of a congruence of timelike geodesics (we will assume the twist is zero):
\begin{equation}
    \frac{d\theta}{d\tau} = - R_{ab}u^au^b - \frac{1}{3}\theta^2 -\sigma_{ab}\sigma^{ab} \label{timelikeexpansion}
\end{equation}
Similarly, the shear evolves according to
\begin{equation}
    \frac{D\sigma_{ab}}{d\tau} = C_{cbad}u^cu^d+\frac{1}{2}\left(\gamma_{ac}\gamma_{bd}-\frac{1}{3}\gamma_{ab}\gamma_{cd}\right)R^{cd}-\frac{2}{3}\theta \sigma_{ab}-\sigma_{ac}\sigma^{c}_{\,b}+\frac{1}{3}\gamma_{ab}\sigma_{cd}\sigma^{cd} \label{timelikeshear}
\end{equation}
where $\gamma_{ab}=g_{ab}+u_au_b$ is the spatial metric orthogonal to the timelike vector field $u^a$. For null geodesics with affine parameter $\lambda$, the analogous evolution equations (with zero twist) are
\begin{align}
    \frac{d}{d\lambda}\hat{\theta} &= - R_{ab}K^aK^b-2\hat{\sigma}^2-\frac{1}{2}\hat{\theta}^2 \label{nullexpansion} \\
    \frac{d}{d\lambda}\hat{\sigma}_{ab} &= - C_{acbd}K^cK^d-\hat{\theta}\hat{\sigma}_{ab}-\hat{\sigma}_{ac}\hat{\sigma}^{c}_b+\frac{1}{2}q_{ab}\hat{\sigma}^{cd}\hat{\sigma}_{cd} \label{nullshear}
\end{align}
where $K^a$ is the tangent vector field of the null congruence and hatted quantities are the null analog defined relative to the 2D surface (with metric $q_{ab}$) orthogonal to the congruence (see, e.g. chapter 9 of \cite{Wald:1984}). We will study the implications of these equations for the homogeneous and isotropic perfect fluid and the planar gravitational shockwave.

\subsubsection{Perfect Fluid}

Conservation of stress-energy for a homogeneous and isotropic perfect fluid gives
\begin{equation}
    \frac{d\rho}{d\tau} = -(1+w)\rho \theta \label{fluidconservation}
\end{equation}
A spacetime consisting of a homogeneous and isotropic perfect fluid is conformally flat, and therefore has no Weyl curvature. The expansion and shear then evolve as
\begin{equation}
    \frac{d\theta}{d\tau} = -4\pi\rho(1+3w)-\frac{1}{3}\theta^2-\sigma_{ab}\sigma^{ab} \label{fluidexpansion}
\end{equation}
\begin{equation}
    \frac{D\sigma_{ab}}{d\tau} = -\frac{2}{3}\theta \sigma_{ab} -\sigma_{ac}\sigma^c_{\,b}+\frac{1}{3}\gamma_{ab}\sigma_{cd}\sigma^{cd} \label{fluidshear}
\end{equation}
In the strictly linear limit, \cref{fluidconservation,fluidexpansion,fluidshear} reduce to
\begin{equation}
    \frac{d\rho}{d\tau} = 0 \label{linfluidconservation}
\end{equation}
\begin{equation}
    \frac{d\theta}{d\tau} = -4\pi \rho (1+3w) \label{linfluidexpansion}
\end{equation}
\begin{equation}
    \frac{d\sigma_{ab}}{d\tau} = 0 \label{linfluidshear}
\end{equation}
In physical language, \cref{linfluidconservation,linfluidexpansion,linfluidshear} say that the density remains constant in the strictly linear regime, the rate of change of expansion is a negative constant, and the shear is constant. The strictly linear model evidently does not consider the mutual gravity of the fluid attracting itself, since we know that an initially static ball of fluid should collapse on itself. In the strictly linear limit, even if the initial expansion of geodesics is positive, since the density remains constant, the expansion will eventually change sign, and the expansion will become increasingly negative. The non-linear solutions remedy this issue by allowing the density to be a decreasing function of time (for initial positive initial expansions), so that $d\theta/d\tau$ and $d\rho/d\tau$ asymptote to zero. 

One in principle could build up an exact solution by solving for the perturbed metric iteratively order by order. Consider a metric perturbation 
\begin{equation}
    g_{ab}=\eta_{ab}+\varepsilon h_{ab}
\end{equation}
Up to $\mathcal{O}(\varepsilon^2)$, the Ricci curvature is given by
\begin{align}
    R_{ab } = &\varepsilon ( \tfrac{1}{2} \nabla_{c }\nabla^{c}h_{ab} + \tfrac{1}{2} \nabla_{a }\nabla_{c }h_{b }{}^{c } + \tfrac{1}{2} \nabla_{b }\nabla_{c }h_{a }{}^{c } -  \tfrac{1}{2} \nabla_{b }\nabla_{a }h) \nonumber \\
    +& \varepsilon^2 ( -  \tfrac{1}{4} \nabla_{\alpha }h^{\beta }{}_{\beta } \nabla^{\alpha }h_{\mu \nu } + \tfrac{1}{2} \nabla^{\alpha }h_{\mu \nu } \nabla_{\beta }h_{\alpha }{}^{\beta } + \tfrac{1}{2} h^{\alpha \beta } \nabla_{\beta }\nabla_{\alpha }h_{\mu \nu } -  \tfrac{1}{2} \nabla_{\alpha }h_{\nu \beta } \nabla^{\beta }h_{\mu }{}^{\alpha } 
    + \tfrac{1}{2} \nabla_{\beta }h_{\nu \alpha } \nabla^{\beta }h_{\mu }{}^{\alpha } \nonumber \\
    &+ \tfrac{1}{4} \nabla_{\alpha }h^{\beta }{}_{\beta } \nabla_{\mu }h_{\nu }{}^{\alpha } -  \tfrac{1}{2} \nabla_{\beta }h_{\alpha }{}^{\beta } \nabla_{\mu }h_{\nu }{}^{\alpha } -  \tfrac{1}{2} h^{\alpha \beta } \nabla_{\mu }\nabla_{\beta }h_{\nu \alpha }+ \tfrac{1}{4} \nabla_{\mu }h^{\alpha \beta } \nabla_{\nu }h_{\alpha \beta } + \tfrac{1}{4} \nabla_{\alpha }h^{\beta }{}_{\beta } \nabla_{\nu }h_{\mu }{}^{\alpha } \nonumber \\
    &-  \tfrac{1}{2} \nabla_{\beta }h_{\alpha }{}^{\beta } \nabla_{\nu }h_{\mu }{}^{\alpha } -  \tfrac{1}{2} h^{\alpha \beta } \nabla_{\nu }\nabla_{\beta }h_{\mu \alpha } + \tfrac{1}{2} h^{c d} \nabla_{b }\nabla_{a }h_{c d}) +\mathcal{O}(\varepsilon^3)\label{perturbedRiccitens}
\end{align}
\begin{align}
    R = &\varepsilon ( \nabla_{\beta }\nabla_{\alpha }h^{\alpha \beta } -  \nabla_{\beta }\nabla^{\beta }h^{\alpha }{}_{\alpha }) \nonumber\\
    + & \varepsilon^2 ( h^{\alpha \beta } \nabla_{\beta }\nabla_{\alpha }h^{\gamma }{}_{\gamma } -  \tfrac{1}{4} \nabla_{\beta }h^{\gamma }{}_{\gamma } \nabla^{\beta }h^{\alpha }{}_{\alpha } -  \nabla_{\alpha }h^{\alpha \beta } \nabla_{\gamma }h_{\beta }{}^{\gamma } + \nabla^{\beta }h^{\alpha }{}_{\alpha } \nabla_{\gamma }h_{\beta }{}^{\gamma } \nonumber\\
    &- 2 h^{\alpha \beta } \nabla_{\gamma }\nabla_{\beta }h_{\alpha }{}^{\gamma } + h^{\alpha \beta } \nabla_{\gamma }\nabla^{\gamma }h_{\alpha \beta } -  \tfrac{1}{2} \nabla_{\beta }h_{\alpha \gamma } \nabla^{\gamma }h^{\alpha \beta } + \tfrac{3}{4} \nabla_{\gamma }h_{\alpha \beta } \nabla^{\gamma }h^{\alpha \beta }) +\mathcal{O}(\varepsilon^3)\label{perturbedRicciscal}
\end{align}
The first order stress-energy tensor can be used to determine the linearized metric perturbation, which then can be fed back into conservation of stress-energy at 2nd order. The 2nd order stress-energy tensor can then be used to determine the the 2nd order metric perturbation, etc. 

For simplicity,  assume the initial shear is zero. The shear then remains zero in the model we are considering. At linear order, we can solve for the expansion and metric perturbation assuming $\rho$ is constant, then plug the (now decreasing) expansion back into \cref{fluidconservation}. One can then solve for the $\mathcal{O}(\varepsilon^2)$ metric perturbation, determine the new expansion at second order, then feed this back into the conservation law, etc. Doing so will ultimately reproduce the exact expansion and density evolution of the radiation dominated FRW model. We now examine how one arrives at this same result by building up the solution from individual gravitational shockwaves sourced by null point particles.

\subsubsection{Shockwave}

As demonstrated in section IV, after summing over photon locations and propagation directions, the linearized Weyl tensor vanishes, and there is only Ricci curvature due to local energy density of the matter. Since the linearized solution is conformally flat, and the full non-linear FRW solution is conformally flat, it must follow that each order in the perturbation series must be conformally flat as well. We interpret this to mean that at every order, the effects of the gravitational shockwaves sourced by null point particles should average out to zero. The deviations in Weyl curvature due to the discreteness of the photon sources will not add linearly beyond linear order, but we argue that the non-linear contributions are small given that each individual photon contributes negligibly compared to the background perfect fluid, and the deviations from 0 of the Weyl tensor should be small so long as the inter-photon spacing is not large.

For a single photon shockwave, conservation of stress energy at linear order gives
\begin{equation}
    \frac{dE}{d\lambda} = 0 \label{photonconservation}
\end{equation}
If we consider effects beyond linear order, conservation of stress energy requires
\begin{equation}
    \frac{dE}{d\lambda} = -E \nabla^aK_a=-E\hat{\theta} \label{photonenergyloss}
\end{equation}
As expected, if the initial expansion of the congruence is positive, the energy of the individual photons decreases. The expansion and shear for a null geodesic (at linear order) evolve with affine parameter $\lambda$ as
\begin{equation}
    \frac{d\theta}{d\lambda} = -8\pi E \delta(x)\delta(y)\delta(u) (k_aK^a)^2\label{photonexpansion}
\end{equation}
\begin{align}
    \frac{d\sigma_{ab}}{d\lambda} = -\frac{4E}{s^2}\delta(u)&\left[\frac{1}{2}k_{(a}s_{b)}(K^cs_c)(K^dk_d)-(s_a\rightarrow \phi_a) \right. \nonumber\\
    &+ \left.(s_as_b-\phi_a\phi_b)(K^ck_c)^2+k_ak_b((K^cs_c)^2-(K^c\phi_c)^2)\vphantom{\frac{1}{2}} \right] \label{photonshear}
\end{align}
As a concrete example, let us take $K^a=t^a-z^a$ for simplicity, such that $K^as_a=K^a\phi_a=0$ and $K^ak_a=-2$. Then we have
\begin{equation}
    \frac{d\theta}{d\lambda} = -32\pi E \delta(x)\delta(y)\delta(u) \label{photonexpansion2}
\end{equation}
\begin{align}
    \frac{d\sigma_{ab}}{d\lambda} = -\frac{16E}{s^2}\delta(u) (s_as_b-\phi_a\phi_b)\label{photonshear2}
\end{align}
From eq. (\ref{photonexpansion2}) and (\ref{photonenergyloss}) we can see that if a congruence of null geodesics with the same initial energy and positive initial expansion were to pass through a 3D array of several consecutive co-planar, counter propagating photon shockwaves, the photon energy and congruence expansion would both decrease over time. In the linear picture we treat the photon array as fixed and study the behavior of the null congruence, but in the full non-linear picture they would interact with each other simultaneously, producing a uniform redshift and decrease in expansion for all photons in consideration. By then allowing the energy and spacing of the 3D array of photons to change, the energy and expansion of the null congruence would asymptote to zero rather than continually become more and more negative. Again, the shear effects would average out to zero for a perfectly homogeneous and isotropic distribution, and would be small for a uniform discrete array. 

\subsection{Redshift}
Previously, we had considered perturbing the spacetime metric around a flat background. In this section, we consider perturbing the metric around a curved background, specifically the FRW spacetime with equation of state $w=1/3$. We shall denote quantities with a tilde to mean the object in the unperturbed spacetime.
\begin{equation}
    g_{ab}=\tilde{g}_{ab}+h_{ab}
\end{equation}
Expanding the Einstein field equations to first order in the trace reversed metric perturbation $\bar{h}_{ab}$ in Lorenz gauge gives
\begin{equation}
    \tilde{g}^ab\nabla_a\nabla_b \bar{h}_{cd}+2\tilde{R}_{cabd}\bar{h}^{ab} = -16\pi T_{ab}
\end{equation}
The covariant derivatives in the wave operator will generate products of Christoffel symbols and derivatives of Christoffel symbols. These terms, along with the Riemann curvature tensor, all depend on first and second derivatives of the scale factor $a(t)$. We can compare the time scale associated with cosmic expansion to the time/length of evolution of the stress-energy tensor components. If the stress-energy tensor components evolve much faster than the rate of cosmic expansion, we can at first order neglect these terms. 

In a general curved spacetime, the stress-energy tensor for a timelike point particle traveling along worldline $\gamma$ parameterized by proper time $\tau$ takes the form
\begin{equation}
    T^{ab}(x) = m\int_{\gamma} u^au^b\frac{\delta^{(4)}(x,z)}{\sqrt{|g|}}d\tau
\end{equation}
In an FRW background spacetime, the stress-energy tensor of a null point particle whose worldline is parameterized by affine parameter $\lambda$ takes the form
\begin{equation}
    T^{ab} = \frac{p}{a(t)^2}k^ak^b\delta(x)\delta(y)\delta(z-\lambda) \label{pointphotonFRW}
\end{equation}
where $k^a$ is the tangent vector to a null geodesic in the background FRW spacetime and $p$ is the conserved momentum given by translation invariance
\begin{equation}
    p_az^a = p = const
\end{equation}
Conservation of stress-energy gives a photon redshift consistent with the typical FRW cosmological redshift.
\begin{equation}
    E = \int d^3x \sqrt{|g|}T_{ab}u^au^b= \frac{p}{a(t)}
\end{equation}

Because all FRW spacetimes are conformally flat, the trajectories of null particles looks the same in conformal time as they would in Minkowski spacetime. Consequently, one would expect propagation of the gravitational shockwave produced by a null point particle to behave similarly in the conformal picture. The key difference is that the energy of photons is redshifted as they propagate. We therefore conclude that over distances short compared to the Hubble length, the picture should be exactly the same. We further conclude that for photons with an impact parameter at moderate distance (not large compared to the Hubble distance), we can simply apply a redshift to the energy of the photon when accounting for the gravitational effects to leading order. This allows us to treat the ``matter'' part of the curvature non-linearly and the vacuum part of the curvature perturbatively. In the next section we will compute classical fluctuations of the gravitational redshift due to the discreteness of the individual photon shockwaves. 

\section{Fluctuations}
Now that we have demonstrated a physical and mathematical correspondence between the gravity of a photon gas and a homogeneous and isotropic collection of gravitational shock waves sourced by null point particles, we will investigate the  fluctuations from a photon gas that are inherited from the micro structure of the gravitational field due to the individual shocks. More explicitly, we define an angular and temporal spectrum of fluctuations by analyzing the temporal and angular dependence of the redshift experienced by a system of spherically arranged clocks as measured by an observer sitting at the center of the system. We then compute the variance of fluctuations in redshift using the angular spectra. 

\subsection{Angular Spectrum}

We are interested in the angular spectrum of gravitational redshift experienced by a system of synchronized clocks arranged spherically around an observer performing a measurement. We can define a redshift factor $z$ via
\begin{equation}
    1+z = \frac{\lambda ' }{\lambda} = \gamma (1+v {\rm cos}\theta_r)
\end{equation}
where $\theta_r$ is the relative angle between the line of sight of the observer and the direction of motion of the clock, and $\gamma$ is the usual Lorentz boost factor. In the low velocity limit, the redshift is dominated by the longitudinal (radial) motion of the clocks relative to the central observer, so we can approximate to leading order $z\approx v^r$. 

Recall that from the geodesic deviation equation, we have
\begin{equation}
    \frac{dv_a}{dt} = - R_{tatb}D^b
\end{equation}
where 
\begin{equation}
   v_b= u^a\nabla_a D_b = D^a\nabla_a u_b \label{relvel}
\end{equation}
Since we have already determined the relative velocity kick $\Delta v^a$ for point-like and Gaussian particles, we can easily determine what $\Delta u^a$ must be. We then decompose these functions in terms of their series representation in the form of spherical harmonics. In standard notation, the angular pattern of a quantity $\Delta$ on a sphere, such as clock reading, scalar curvature or CMB temperature,  can be decomposed into spherical harmonics
$Y_{\ell m}(\theta,\phi)$:
\begin{equation}\label{decompose}
\Delta(\theta,\phi)=
\sum_\ell \sum_m Y_{\ell m}(\theta,\phi) a_{\ell m}.
\end{equation}
The harmonic coefficients $a_{\ell m}$ then determine
the angular power spectrum:
\begin{equation}\label{powerpiece}
C_\ell= \frac{1}{2\ell+1}
\sum_{m= -\ell}^{m=+\ell} | a_{\ell m}|^2.
\end{equation} 
The angular correlation function is given by its Legendre transform,
\begin{equation}\label{harmonicsum}
 C(\Theta) = \frac{1}{4\pi}\sum_\ell (2\ell +1) { C}_\ell P_\ell (\cos \Theta),
\end{equation}
where $P_\ell$ are Legendre polynomials.

The angular correlation function also corresponds to an all-sky average
\begin{equation}
 C(\Theta)=\langle \Delta_1 \Delta_2\rangle_{\Theta}
\end{equation}
for all pairs of points $1,2$  separated by angle $\Theta$, or equivalently, an  average over all directions $i$
\begin{equation}\label{skyaverage}
 C(\Theta)=\langle \Delta_i \bar\Delta_{i\Theta}\rangle_i
\end{equation}
where $\bar\Delta_{i\Theta}$ denotes the azimuthal average on a circle of angular radius $\Theta$  with center $i$. Azimuthally symmetric bounds on causal relationships from intersections of null surfaces thus map directly onto bounds of $C(\Theta)$, rather than the spectral domain $C_\ell$:
sharp  boundaries of causal relationships produce  sharp boundaries of  angular correlation.

In the current subsection we will focus on studying the $a_{\ell m}$, then in the subsequent subsection we will turn our attention to the $C_{\ell}$ and derive the angular correlation functions. A summary of definitions is provided below.
\begin{equation}
    Y_{\ell m}(\theta,\phi) = \sqrt{\frac{2\ell+1}{4\pi}}\sqrt{\frac{(\ell-m)!}{(\ell+m)!}}P_{\ell}^m({\rm cos}\theta)e^{im\phi} \label{sphericalharmonics}
\end{equation}
\begin{equation}
    P_{\ell}^m(x) = (-1)^m(1-x^2)^{m/2}\frac{d^m}{dx^m}P_{\ell}(x)\label{associatedlegendreP}
\end{equation}
\begin{equation}
    P_{\ell}(x) = \frac{1}{2^{\ell}\ell!}\frac{d^{\ell}}{dx^{\ell}}(x^2-1)^{\ell}\label{legendreP}
\end{equation}

\subsubsection*{i. Small Impact Parameter}
\begin{figure}
    \centering
    \includegraphics[width=.3\linewidth]{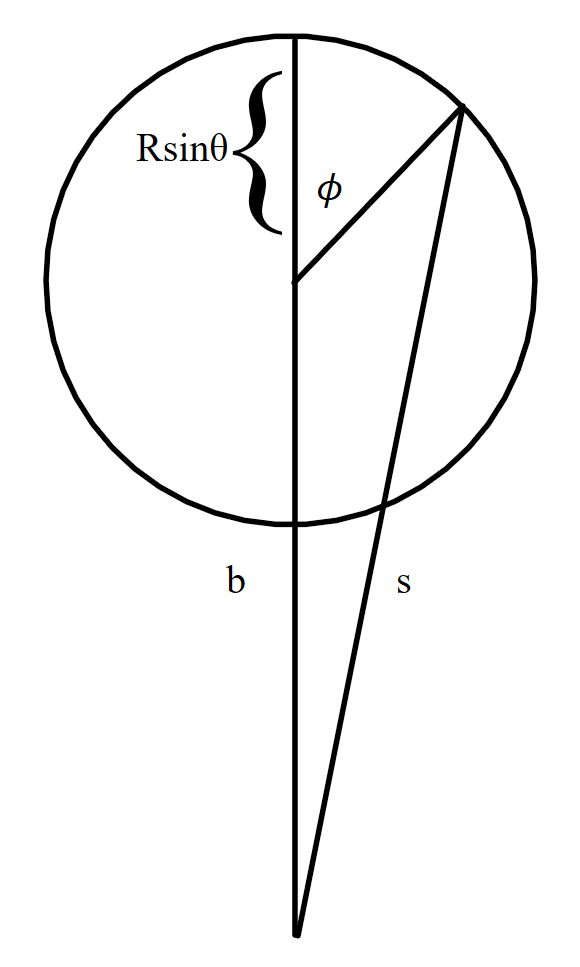}
    \caption{Geometry for finite impact parameter. The gravitational effect of null point particles depends only on the distance from the particle in the transverse plane. Each circular cross-section of the sphere with radius $R$ has radius $R {\rm sin}\theta$.}
    \label{fig:impact_parameter_geom}
\end{figure}

For a small impact parameter, it is best to work with the Gaussian model for the photon since it avoids singularities in the velocity kick at the poles ($\theta=0,\pi$). Using the geodesic deviation equation along with eq. (\ref{relvel}), one can readily determine the velocity kick imparted on a test body due to the Gaussian photon shockwave. 
\begin{equation}
    \Delta u^a = -\frac{4E}{s}\left(1-e^{-s^2/2a^2}\right)s^a \label{gaussiankick}
\end{equation}
This agrees with the velocity kick of a point particle in the far-field limit $s>>a$. Consider a source which is characterized by an impact parameter $b$ measured relative to the center of the sphere in the $(x,y)$ plane (see \cref{fig:impact_parameter_geom}). Measured from the symmetry axis of the source, we have
\begin{equation}
    s^2 = b^2 + R^2{\rm sin}^2\theta + 2 b R {\rm sin}\theta {\rm cos}\phi \label{impactparameter}
\end{equation}
where $(r,\theta,\phi)$ are spherical polar coordinates adapted to an observer in the middle of our system of clocks. Then we have
\begin{equation}
    s^a = {\rm sin}\theta  \left(\frac{R}{s}{\rm sin}\theta +\frac{b}{s}{\rm cos}\phi\right)r^a + {\rm cos}\theta \left(\frac{R}{s}{\rm sin}\theta +\frac{b}{s}{\rm cos}\phi\right)\theta^a-\frac{b}{s}{\rm sin}\theta {\rm sin}\phi \phi^a
\end{equation}

\begin{figure}[H]
    \centering
    \includegraphics[width=.7\linewidth]{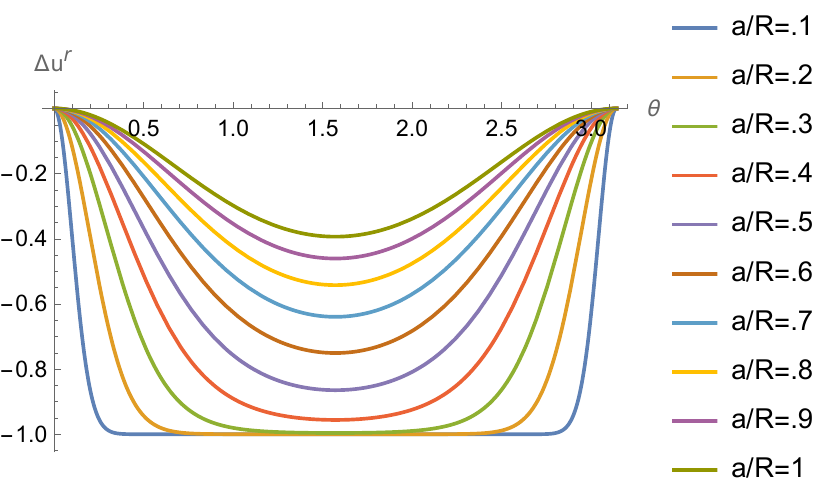}
    \caption{Radial velocity kick profile (\cref{0bgaussiankick}) for a photon with zero impact parameter relative to the observer at the center of a sphere. Units are normalized by a factor of $4E/R$. Localization of the photon decreases from bottom to top in the figure.}
    \label{fig:0b_radial_kick}
\end{figure}

\begin{figure}[t]
    \centering
    \includegraphics[width=.7\linewidth]{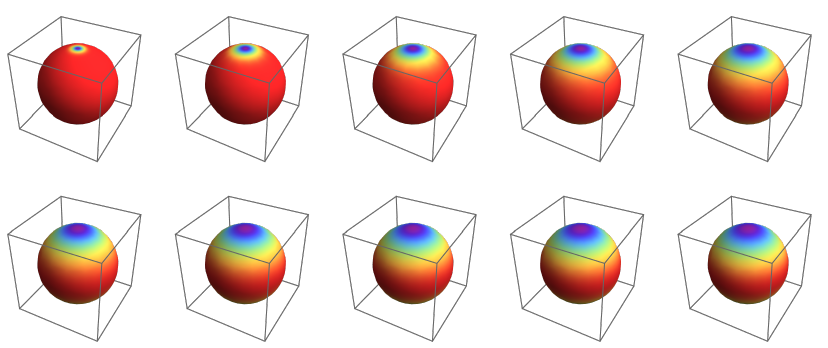}
    \caption{Spherical heat map representing the angular profile of gravitational redshift on the sphere for a particle with zero impact parameter. Figures range from $a/R=.1\rightarrow 1$ moving from left to right, top to bottom.  Red represents the maximal (absolute) value, while purple represents the minimal value. These figures are consistent with the trends shown in \cref{fig:0b_radial_kick}.}
    \label{fig:0b_sphere_heatmap}
\end{figure}

First consider the zero impact parameter case where $r=R$ is where the clocks are initially fixed: 
\begin{equation}
    \Delta u^a = -\frac{4E}{R}\left(1-e^{-R^2{\rm sin}^2\theta/2a^2}\right)(r^a+{\rm cot}\theta \theta^a) \label{0bgaussiankick}
\end{equation}
For zero impact parameter, the redshift angular profile depends only on $\theta$, so we can visualize the profile using a simple 2D plot (shown in \cref{fig:0b_radial_kick}). A highly localized photon (i.e. $a<<R$) has a redshift profile which is nearly constant for a wide range of angles, except very close to the poles (i.e. very close to the photon). The same information is shown in \cref{fig:0b_sphere_heatmap}  using a spherical heat map, which normalizes the amplitudes automatically and focuses on the relative values as a function of angle on the sphere. 

For a a highly de-localized photon (i.e. $a>>R$), eq. (\ref{0bgaussiankick}) approximates to 
\begin{equation}
    \Delta u^a \approx -\frac{2E}{a}\left(\frac{R}{a}\right)({\rm sin}^2\theta r^a + {\rm sin}\theta {\rm cos}\theta \theta^a)
\end{equation}
at leading order. In terms of spherical harmonics, we have
\begin{equation}
    \Delta u^r = -\frac{8 \sqrt{\pi} E}{3a}\left(\frac{R}{a}\right)\left(Y_{0\,0}(\theta,\phi)-\frac{1}{\sqrt{5}}Y_{2\,0}(\theta,\phi)\right) \label{delocalizedkick}
\end{equation}
Additional terms with powers $(R/a)^{n}$ will introduce angular terms scaling like ${\rm sin}^{2n}(\theta)$, and therefore will include terms up to $Y_{2n\, 0}(\theta,\phi)$ in the angular spectrum. Note that the axial symmetry for the zero impact parameter cases requires there to be no non-zero $m$ modes (this will not be the case for finite impact parameter). 

In the highly localized limit ($a<<R$), we must keep all of the terms in the series expansion for the exponential term. Normally one could simply ignore the exponential term for small $a$, except that the ${\rm sin}\theta$ term goes to zero at the poles and the argument of the exponential is no longer large anymore. Recall that
\begin{equation}
    e^x = \sum_{n=0}^{\infty}\frac{x^n}{n!}
\end{equation}
is a convergent series with infinite radius of convergence. Then we have
\begin{align}
    1-e^{-R^2{\rm sin}^2\theta/2a^2} &= 1-\sum_{n=0}^{\infty}\frac{(-1)^n}{n!2^n}\left(\frac{R}{a}\right)^{2n} {\rm sin}^{2n}\theta \\
    &= \sum_{n=1}^{\infty}\frac{(-1)^{n+1}}{n!2^n}\left(\frac{R}{a}\right)^{2n} {\rm sin}^{2n}\theta
\end{align}
The goal is to find an angular spectrum for this expression in terms of spherical harmonics. Since there is no $\phi$ dependence, all terms will have $m=0$. In other words, we seek an expression of the form\footnote{The cutoff at $\ell=2n$ is due to the fact that the $Y_{\ell m}$ include terms only up to ${\rm cos}^{2n}\theta$. One can also see this when using integration by parts $\ell$ times in eq. (\ref{xbasisintegral})}
\begin{equation}
   \Delta u^r = -\frac{4E}{R} \sum_{n=1}^{\infty}\sum_{\ell=0}^{2n} \left(\frac{R}{a}\right)^{2n} A_{n,\ell} Y_{\ell \,0}(\theta,\phi) \label{0bseriesexpansion}
\end{equation}
Using the notation from above, we have (up to an overall factor of $-4E/R$)
\begin{equation} \label{0bharmoniccoeff}
    a_{\ell \,m} = \sum_{n=1}^{\infty}\left(\frac{R}{a}\right)^{2n}A_{n\ell} \quad (\ell \, \rm{even})
\end{equation}
In the sum above, any coefficient $A_{n,\ell}$ such that $n<\ell/2$ is identically zero. 
Since the spherical harmonics form an orthonormal basis of functions on the sphere, we have that
\begin{equation}
    A_{n,\ell} = \frac{(-1)^{n+1}}{n!2^n}B_{n,\ell}
\end{equation}
where
\begin{equation}
    B_{n,\ell} = \int d\Omega \,{\rm sin}^{2n}\theta \,Y_{\ell \,0}(\theta,\phi)
\end{equation}
To evaluate these integrals, it is simplest to make the usual variable change $x={\rm cos}\theta$. Using the binomial theorem,
\begin{equation}
    (1+y)^n= \sum_{k=0}^n\frac{n!}{k!(n-k)!}y^{n-k} \quad n\in \mathbb{Z}
\end{equation}
and recalling the definition of Legendre Polynomials from \cref{legendreP}, we find that in order to determine the $B_{n,\ell}$, we need to evaluate integrals of the form
\begin{equation}
    \int_{-1}^1 dx (x^2-1)^n \frac{d^{\ell}}{dx^{\ell}}(x^2-1)^{\ell} \label{xbasisintegral}
\end{equation}
The final result is 
\begin{align} \label{0bangularcoeff}
    A_{n,\ell} =& \frac{(-1)^{n+1}\sqrt{\pi(2\ell+1)}}{2^{\ell+n-1}} \times \\
    &\sum_{k=0}^n\sum_{p=0}^{\ell/2}\frac{(-1)^{k+p}}{k!(n-k)!p!(\ell-p)!}\frac{(2\ell-2p)!}{(\ell-2p)!}\frac{1}{2k+\ell-2p+1} \quad (\ell \; even) \nonumber
\end{align}
As a quick check, we find that eq. (\ref{0bseriesexpansion}) keeping only the leading order terms $A_{1,0},A_{1,2}$ gives the expression in eq. (\ref{delocalizedkick}). Further, one can readily verify that $|A_{n+1,\ell}/A_{n,\ell}|$ tends to zero rapidly as well as that $|A_{n,\ell+1}|<|A_{n,\ell}| \; \forall\; \{n |\ell\leq 2n\}$. 

The result for arbitrary impact parameter is quite complex, so we will focus our attention only on the leading order behavior for small impact parameter. 
\begin{align}
    \Delta u^r &\approx  -\frac{4E}{R}\left(1-e^{-R^2{\rm sin}^2\theta/2a^2}\right) \label{smallbkick}\\
    &+\frac{4Eb}{R^2}{\rm csc}\theta {\rm cos}\phi\left(1-e^{-R^2{\rm sin}^2\theta/2a^2}\left(1+\frac{R^2{\rm sin}^2\theta}{a^2}\right)\right) \nonumber
\end{align}
Evaluating eq. (\ref{smallbkick}) at $R=0$ gives
\begin{equation}
    \Delta u^r (R=0) \approx -\frac{2Eb}{a^2}{\rm sin}\theta {\rm cos}\phi
\end{equation}
for the radial velocity kick experienced by the observer. Subtracting this from eq. (\ref{smallbkick}) we find a relative kick between the clocks and the observer given by
\begin{align}
    \Delta u^r \approx & -\frac{4E}{R}\left(1-e^{-R^2{\rm sin}^2\theta/2a^2}\right) \label{smallbrelativekick} \\
    &+\frac{2Eb}{a^2}{\rm sin}\theta {\rm cos}\phi\left(\frac{2a^2}{R^2{\rm sin}^2\theta}\left(1-e^{-R^2{\rm sin}^2\theta/2a^2}\right)+\left(1-2e^{-R^2{\rm sin}^2\theta/2a^2}\right)\right) \nonumber
\end{align}
One can derive a spectrum for the term proportional to $b$ in eq. (\ref{smallbrelativekick}). The key take away is that the spectrum will now include the odd $\ell$ harmonics, in addition to the $m=1$ azimuthal term. More generally, one can show that higher order corrections proportional to $b^n$ will have odd $\ell$ harmonics if $n$ is odd and even $\ell$ harmonics if $n$ is even. In additional, each term will contain terms in the azimuthal spectrum up to $m=\pm n$. It is also worth mentioning that all terms proportional to $b^n$ integrate to zero when averaging over the sphere. 

\begin{figure}[H]
    \centering
    \includegraphics[width=.7\linewidth]{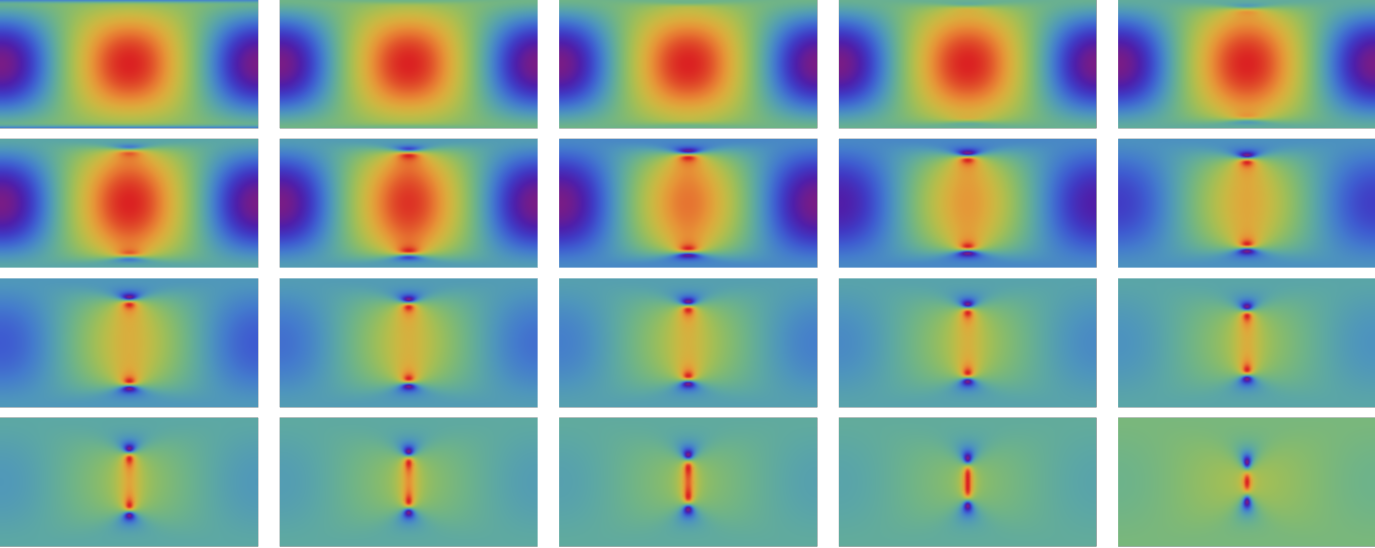}
    \caption{Series of heat maps representing the (normalized) radial velocity kick/redshift \cref{smallbrelativekick} for impact parameter $0<b<R$, $a/R=.05$. Horizontal and vertical axes are the spherical coordinates $(\phi,\theta)$ respectively. Impact parameter increases going from left to right, top to bottom. Spectrum is mostly dipolar for small impact parameter. }
    \label{fig:smallbheatmap}
\end{figure}

\begin{figure}[t]
    \centering
    \includegraphics[width=.75\linewidth]{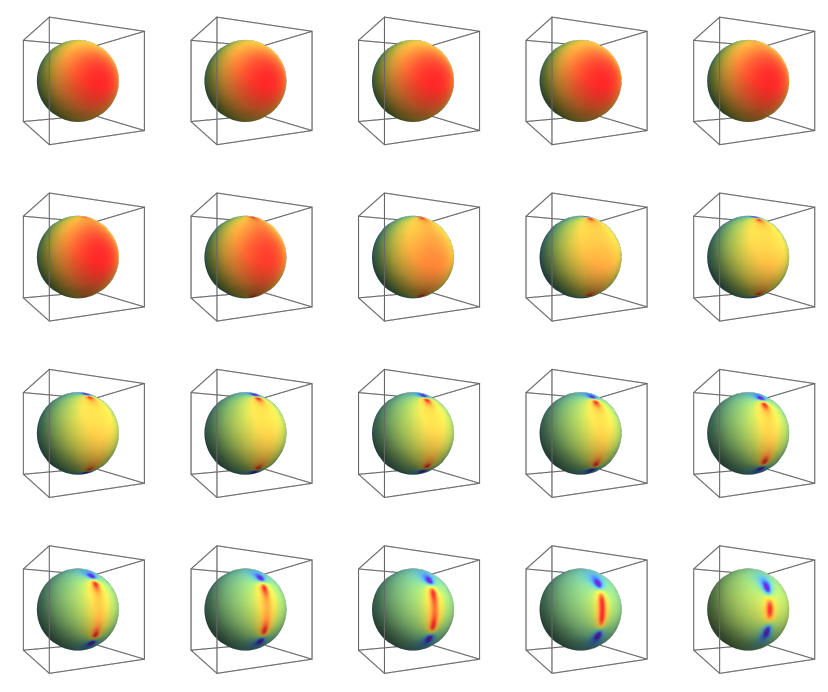}
    \caption{Heat maps from \cref{fig:smallbheatmap} represented on the sphere. Impact parameter increases going from left to right, top to bottom.}
    \label{fig:smallbsphereheatmap}
\end{figure}

The angular spectrum for the redshift can be best visualized using spherical heat maps. \cref{fig:smallbheatmap,fig:smallbsphereheatmap} shows a series of heat maps for $a/R=.05$, and $0<b<R$. For small but non-zero impact parameter, the angular profile is mostly dipolar, and predominantly due to the kick experienced by the observer. For $b\approx R$, all multipole moments must be considered.

\subsubsection*{ii. Large Impact Parameter}
 In the limit of large impact parameter $b>>R,a$, the Gaussian term in eq. (\ref{gaussiankick}) becomes negligible. The radial velocity kick is then well approximated by
 \begin{equation}
   \Delta u^r = -\frac{4E}{s^2}{\rm sin}\theta(R {\rm sin}\theta + b {\rm cos}\phi) \label{largebkick}
\end{equation}
Recall that for $|x|<1$, we have
\begin{equation}
    \frac{1}{1+x} = \sum_{n=0}^{\infty}(-1)^n x^n
\end{equation}
Therefore we can perform a series expansion in $y=R{\rm sin}\theta/b<<1$ such that
\begin{equation}
    \frac{1}{s^2} = \frac{1}{b^2}\sum_{n=0}^{\infty}(-1)^n \left(y^2+2 y {\rm cos}\phi\right)^n
\end{equation}
The result for eq. (\ref{largebkick}) simplifies to
\begin{align}
    \Delta u^r &= -\frac{4E}{b}\sum_{\ell=1}^{\infty} (-1)^{\ell+1} {\rm cos}(\ell \phi) {\rm sin}^{\ell}\theta \left(\frac{R}{b}\right)^{\ell-1} \\
    &= \frac{-4E}{b}\sum_{\ell=1}^{\infty}(-1)^{\ell+1}2^{\ell-1}\ell!\sqrt{\frac{4\pi}{(2\ell+1)!}}((-1)^{\ell}Y_{\ell \,\ell}(\theta,\phi)+Y_{\ell \,-\ell}(\theta,\phi))\left(\frac{R}{b}\right)^{\ell-1} \label{largebkickseries}
\end{align}
\begin{figure}[H]
    \centering
    \includegraphics[width=.7\linewidth]{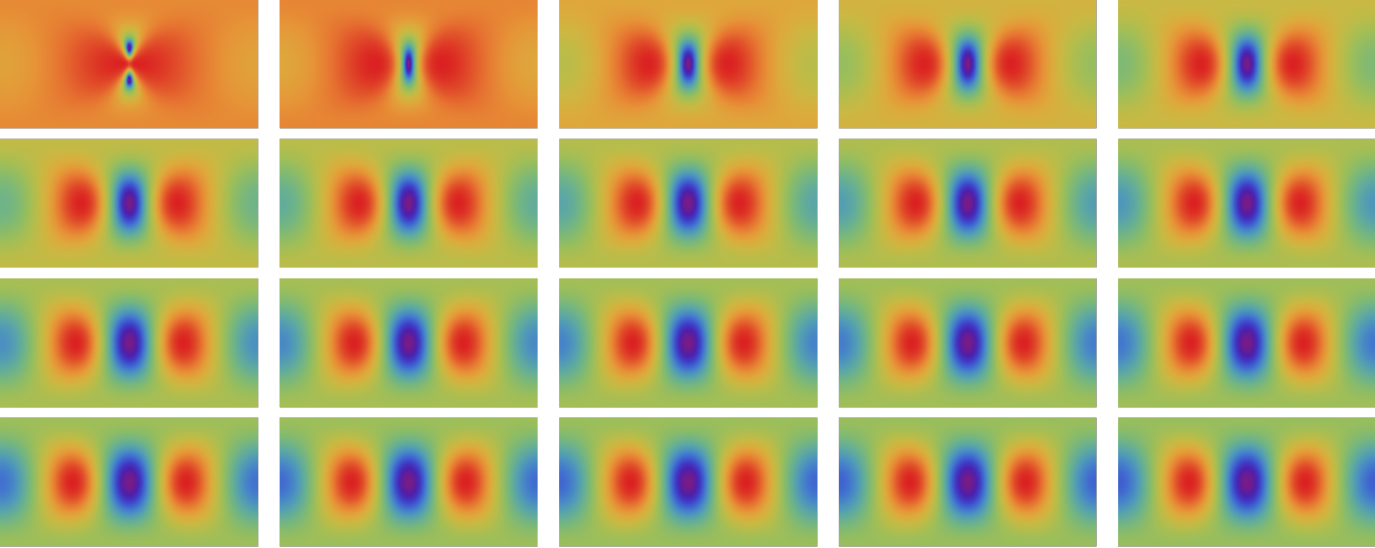}
    \caption{Series of heat maps for radial velocity kick/redshift (\cref{largebkickseries}) for impact parameters $R\leq b \leq 2R$ ($a/R=.05$). The spectrum quickly asymptotes to a predominantly quadrupolar profile. Impact parameter increases moving from left to right, top to bottom.}
    \label{fig:largebheatmap}
\end{figure}
\begin{figure}[t]
    \centering
    \includegraphics[width=.75\linewidth]{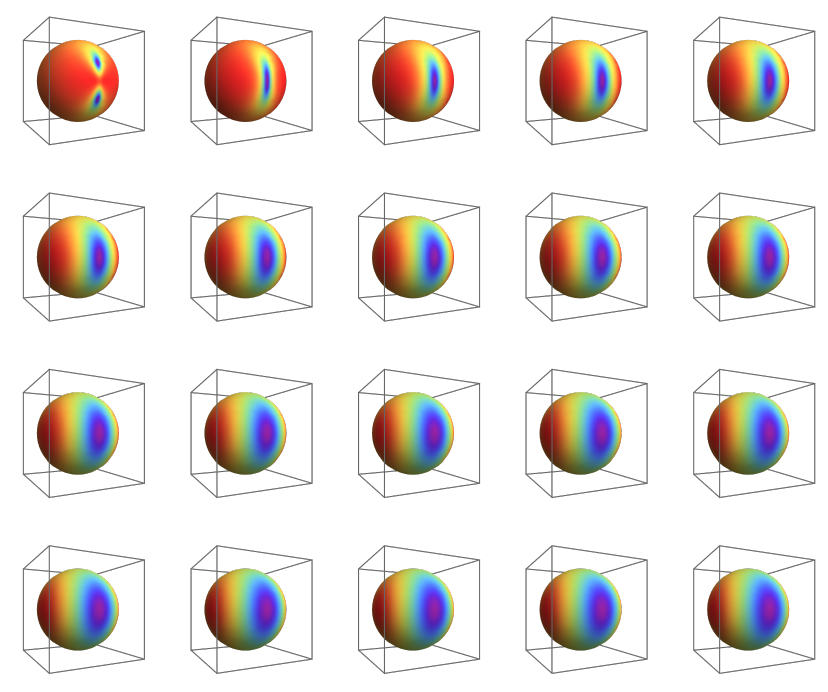}
    \caption{Heat maps from \cref{fig:largebheatmap} represented on the sphere. Impact paremeter increases from left to right, top to bottom.}
    \label{fig:largebsphereheatmap}
\end{figure}

Since we are interested in the radial velocity kick \textit{relative} to the central observer, we need to subtract off the velocity kick experienced by the observer. This amounts to subtracting the value of eq. (\ref{largebkickseries}) evaluated at $R=0$. Since only the $\ell=1$ term is non-zero for $R=0$, this operation is equivalent to eliminating the dipole moment\footnote{This is to be expected since the dipole moment of any stress-energy tensor is simply the momentum, and returning to the center of momentum frame will eliminate any dipole term.}, thus the first contribution is the quadrupole moment.  

\newpage
\subsection{Angular Correlation}
In this section, we turn our attention to the angular correlation functions defined by \cref{powerpiece,harmonicsum}. We provide a series of angular correlation functions for a range of different combinations of the parameters $R,a,b$. 

\subsubsection*{i. Small Impact Parameter}
\begin{figure}[H]
    \centering
    \includegraphics[width=.7\linewidth]{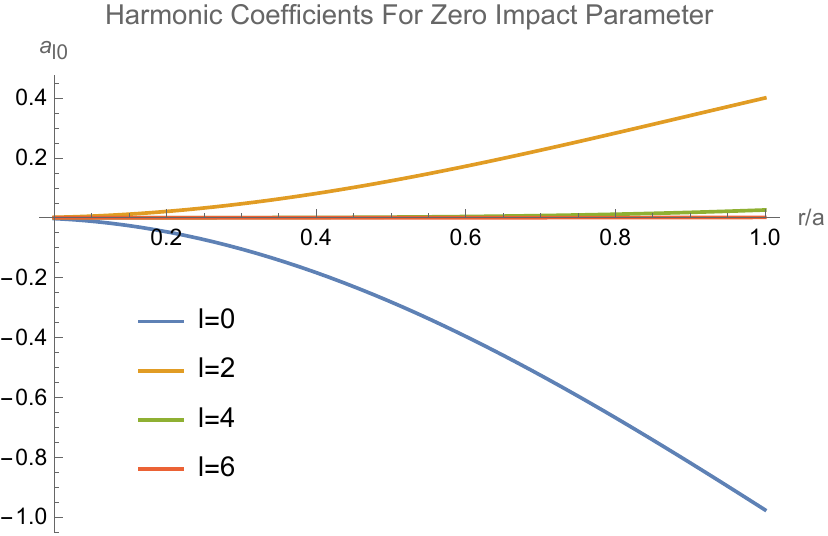}
    \caption{Harmonic coefficients $a_{\ell \,0}$ (\cref{0bharmoniccoeff2}) as a function of $r/a$ ($b=0$). For a delocalized photon ($a>R$), the spectrum is dominated by the monopole and quadrupole terms. }
    \label{fig:zerobharmoniclargea}
\end{figure}

\begin{figure}[H]
    \centering
    \includegraphics[width=.7\linewidth]{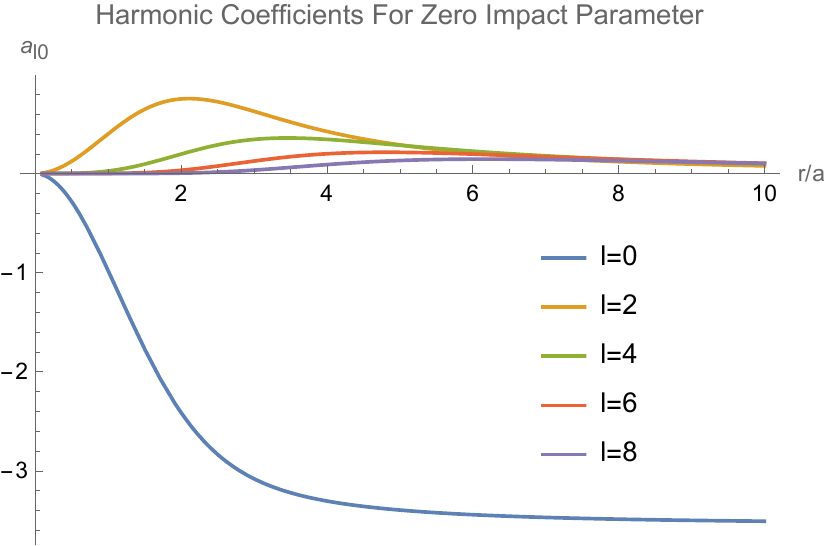}
    \caption{Harmonic coefficients $a_{\ell \,0}$ (\cref{0bharmoniccoeff2}) as a function of $r/a$ ($b=0$). For highly localized photons ($a<<R$), the spectrum is dominated by the monopole.} 
    \label{fig:zerobharmonicsmalla}
\end{figure}

\begin{figure}[H]
    \centering
    \includegraphics[width=.7\linewidth]{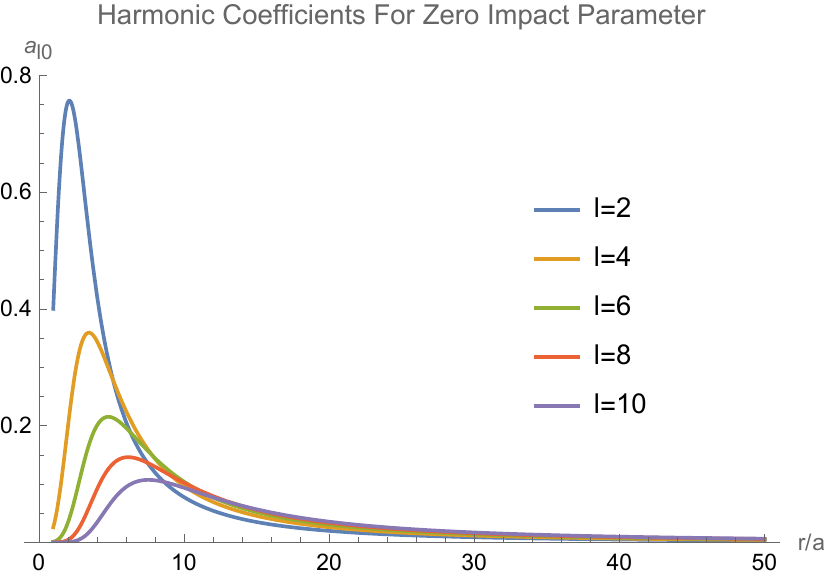}
    \caption{Harmonic coefficients $a_{\ell \,0}$ (\cref{0bharmoniccoeff2}) as a function of $r/a$ ($b=0$). With the monopole part removed, all other multipole moments scale as $\mathcal{O}(a/R)^2$ in the asymptotic limit $a\rightarrow 0$. The multipole moment expansion therefore breaks down in this limit, as all $\ell$ values contribute. }
    \label{fig:zerobharmonicsmalla2}
\end{figure}

\begin{figure}[H]
    \centering
    \includegraphics[width=.7\linewidth]{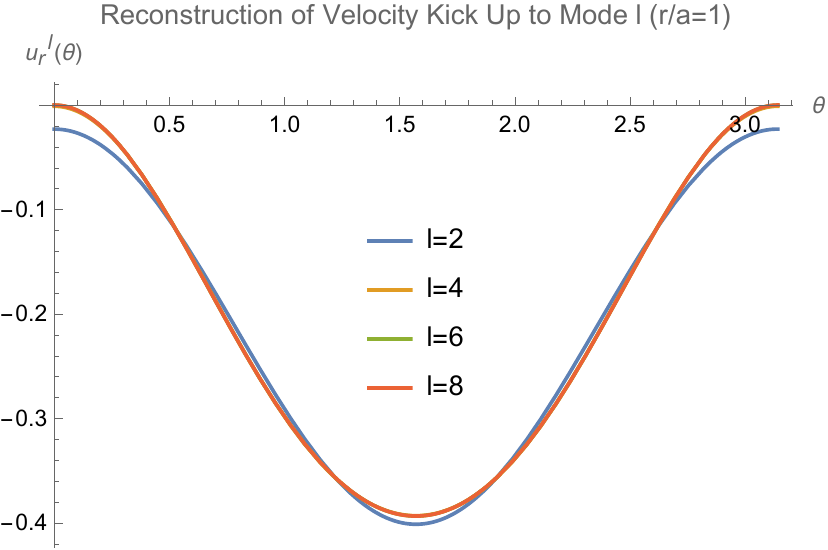}
    \caption{Reconstruction of radial velocity kick profile up to mode $\ell$ for delocalized photon $R=a,b=0$. Only a small number of modes is needed to accurately represent the full function.}
    \label{fig:0bvelkickreconlargea}
\end{figure}

\begin{figure}[H]
    \centering
    \includegraphics[width=.7\linewidth]{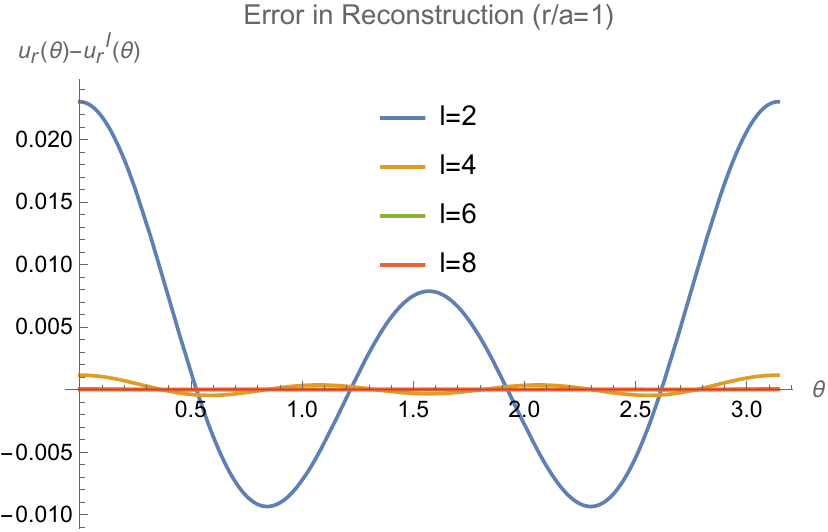}
    \caption{Error in the radial velocity kick reconstruction reconstruction up to mode $\ell$ for $R=a,b=0$. As mentioned in \cref{fig:0bvelkickreconlargea}, the error can be made arbitrarily small with only a small number of modes. }
    \label{fig:0bvelreconerrorlargea}
\end{figure}

\begin{figure}[H]
    \centering
    \includegraphics[width=.7\linewidth]{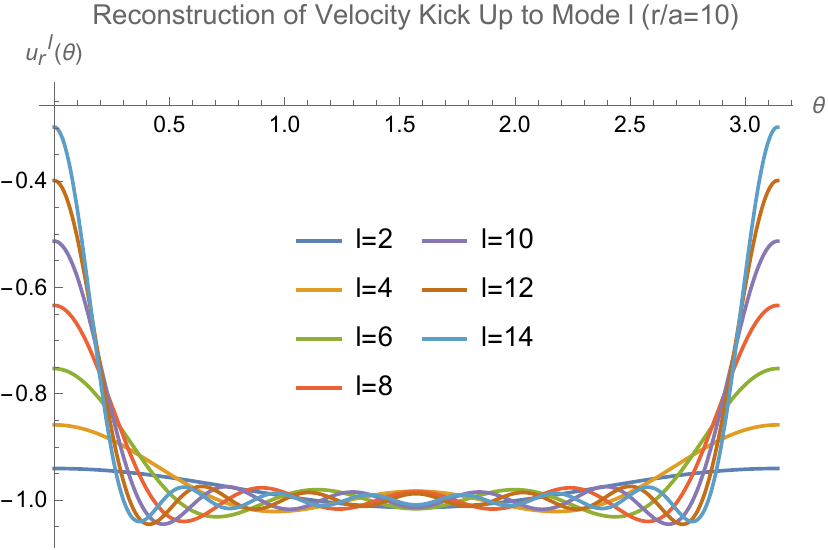}
    \caption{Reconstruction of radial velocity kick profile up to mode $\ell$ for $a=.1R,b=0$. As $a$ tends to zero, the number of $\ell$ modes need to accurately capture the behavior tends to infinity.}
    \label{fig:0bvelkickreconsmalla}
\end{figure}

\begin{figure}[H]
    \centering
    \includegraphics[width=.7\linewidth]{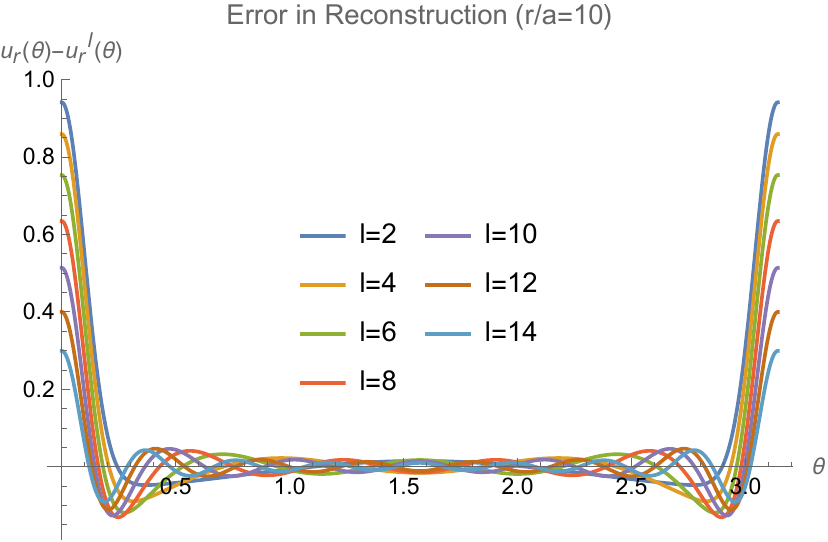}
    \caption{Error in radial velocity kick reconstruction up to mode $\ell$ for $a=.1R,b=0$. Error is small away from poles for all $\ell$, but is very large near the poles when only a small number of $\ell$ modes is used.}
    \label{fig:0bvelkickreconerrorsmalla}
\end{figure}

\begin{figure}[H]
    \centering
    \includegraphics[width=.7\linewidth]{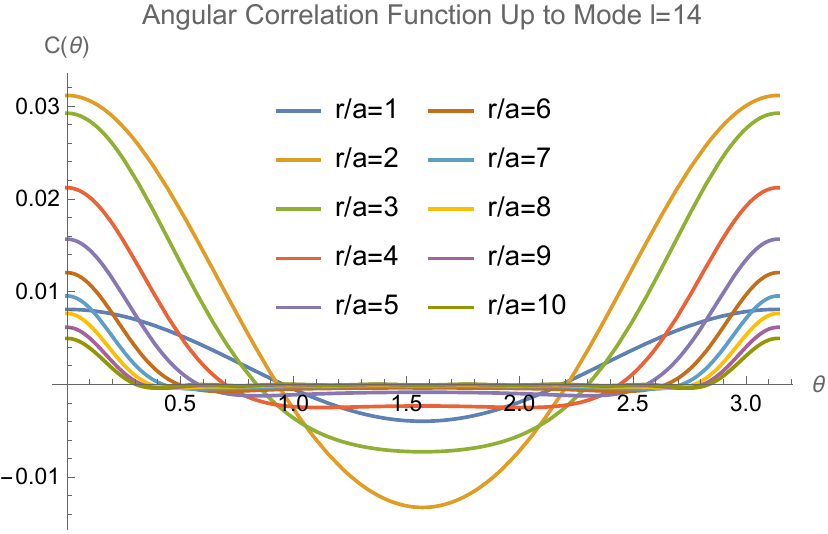}
    \caption{Plot of angular correlation function $C(\Theta)$ ($b=0$) summed up to mode $\ell=14$ (monopole subtracted). For $a\sim R$, the spectrum is primarily quadrupolar. For $a<<R$, the spectrum picks up higher multipole moments. Note the wide angular range of nearly-vanishing correlation, a signature of causal symmetry not apparent from  power spectra alone. }
    \label{fig:0bangcorr}
\end{figure}

\begin{figure}[H]
    \centering
    \includegraphics[width=.7\linewidth]{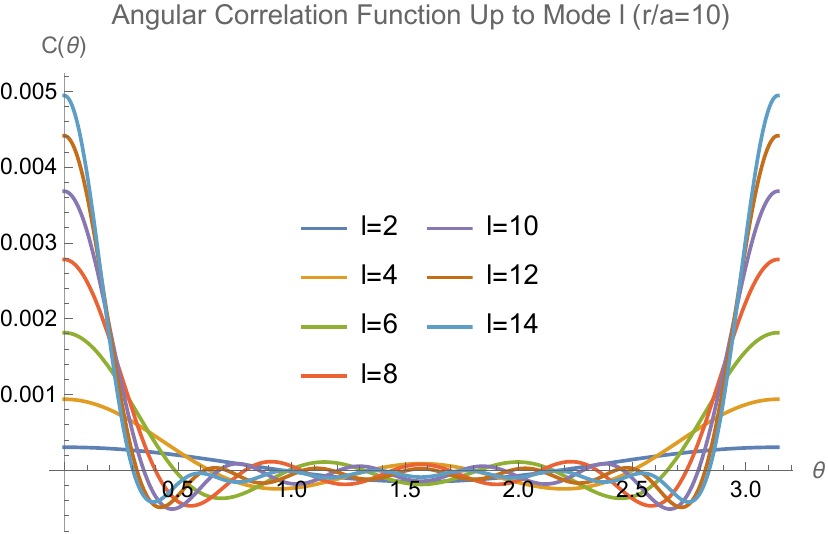}
    \caption{Plot of angular correlation function $C(\Theta)$ summed up to mode $\ell$ for $a=.1R,b=0$. }
    \label{fig:0bangcorr2}
\end{figure}

We first consider the case of zero impact parameter. We can determine an exact power series representation for the harmonic coefficients $a_{\ell,0}$ using \cref{0bharmoniccoeff,0bangularcoeff}. The general result is
\begin{equation}
    a_{\ell,0} = \frac{1}{(r/a)^{\ell+1}}(f_{\ell-1}(r/a)+g_{\ell}(r/a)D_+(r/\sqrt{2}a)) \label{0bharmoniccoeff2}
\end{equation}
where $D_+(x)$ is the Dawson exponential integral (defined in detail in section VI C.), and $f_{\ell-1}(x),g_{\ell}(x)$ are generic polynomials in $x$ up to order $\ell-1$ and $\ell$, respectively. The various harmonics are shown in \cref{fig:zerobharmoniclargea,fig:zerobharmonicsmalla,fig:zerobharmonicsmalla2}. The angular correlation functions are shown in \cref{fig:0bangcorr,fig:0bangcorr2}. The reconstruction of the velocity kick is shown in \cref{fig:0bvelkickreconlargea,fig:0bvelreconerrorlargea,fig:0bvelkickreconsmalla,fig:0bvelkickreconsmalla,fig:0bvelkickreconerrorsmalla}. For a highly localized photon, many angular modes are needed to reproduce the angular spectrum. For a delocalized photon, the monopole and quadrupole terms dominate. 


\begin{figure}[H]
    \centering
    \includegraphics[width=.7\linewidth]{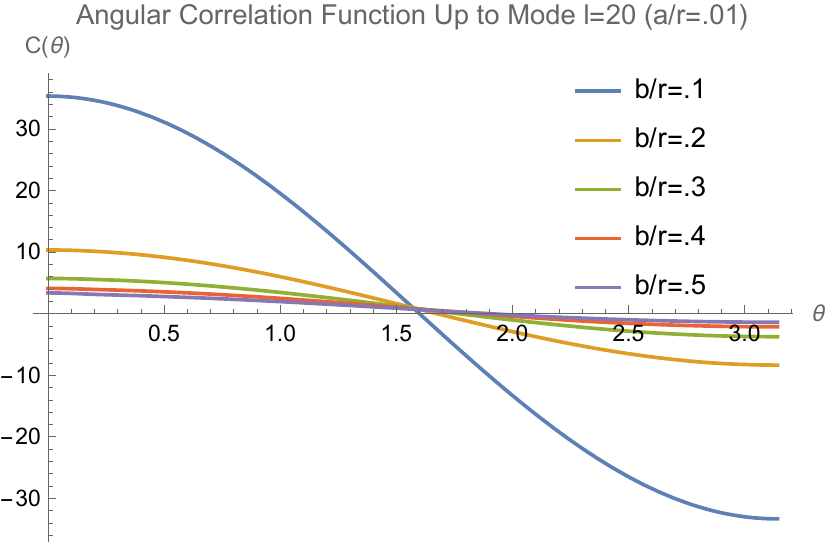}
    \caption{Angular correlation function $C(\Theta)$ for small impact parameter and highly localized photon $(a=.01R)$. Spectrum is predominantly dipolar.}
    \label{fig:smallbsmallaangcorr}
\end{figure}

\begin{figure}[H]
    \centering
    \includegraphics[width=.7\linewidth]{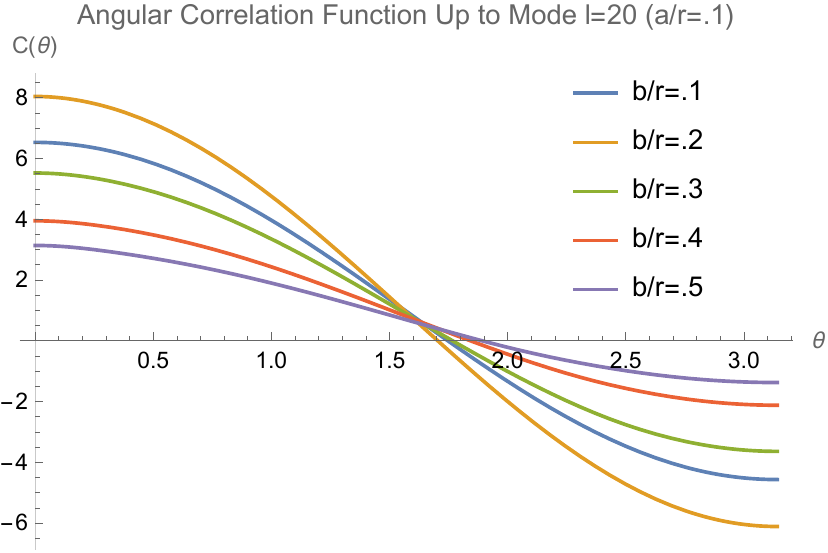}
    \caption{Angular correlation function $C(\Theta)$ for small impact parameter and somewhat localized photon $(a=.1R)$. Spectrum is still predominantly dipolar but picks up some monopole contributions. }
    \label{fig:smallbmidaangcorr}
\end{figure}

\begin{figure}[H]
    \centering
    \includegraphics[width=.7\linewidth]{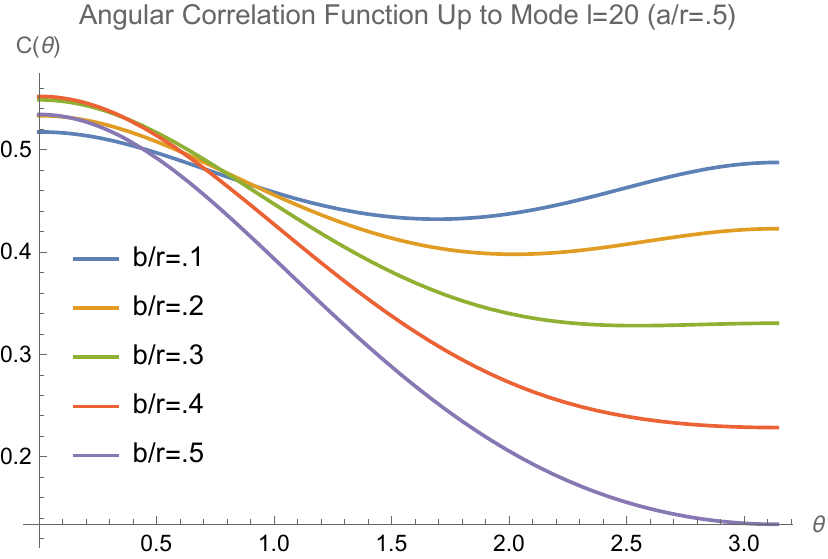}
    \caption{Angular correlation function $C(\Theta)$ for small impact parameter and  delocalized photon $(a=.5R)$.}
    \label{fig:smallblargeaangcorr}
\end{figure}

\begin{figure}[H]
    \centering
    \includegraphics[width=.7\linewidth]{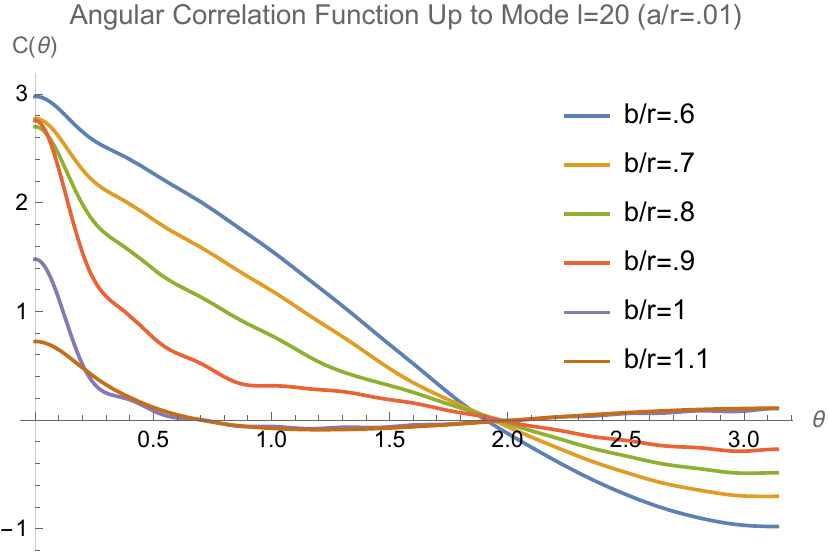}
    \caption{Angular correlation function $C(\Theta)$ for $b\sim R$ and highly localized photon $(a=.01R)$.}
    \label{fig:midbsmallaangcorr}
\end{figure}

\begin{figure}[H]
    \centering
    \includegraphics[width=.7\linewidth]{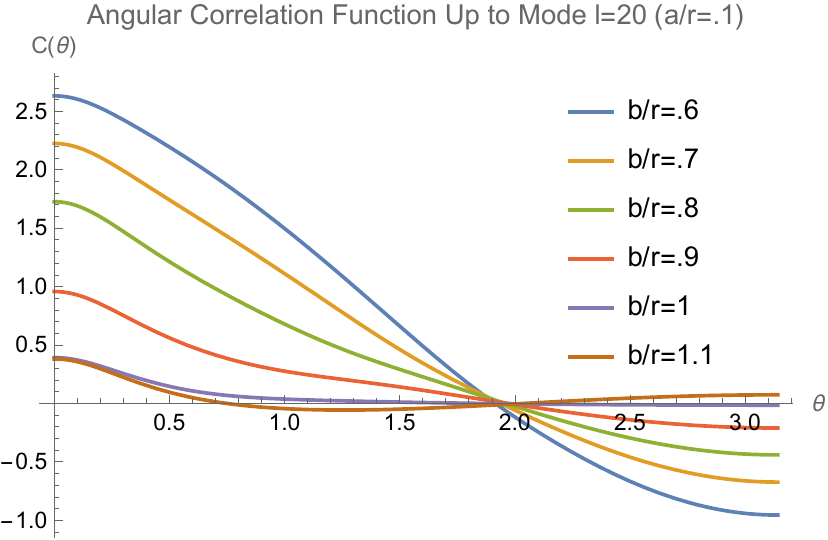}
    \caption{Angular correlation function $C(\Theta)$ for $b\sim R$ and somewhat localized photon $(a=.1R)$.  }
    \label{fig:midbmidaangcorr}
\end{figure}

\begin{figure}[H]
    \centering
    \includegraphics[width=.7\linewidth]{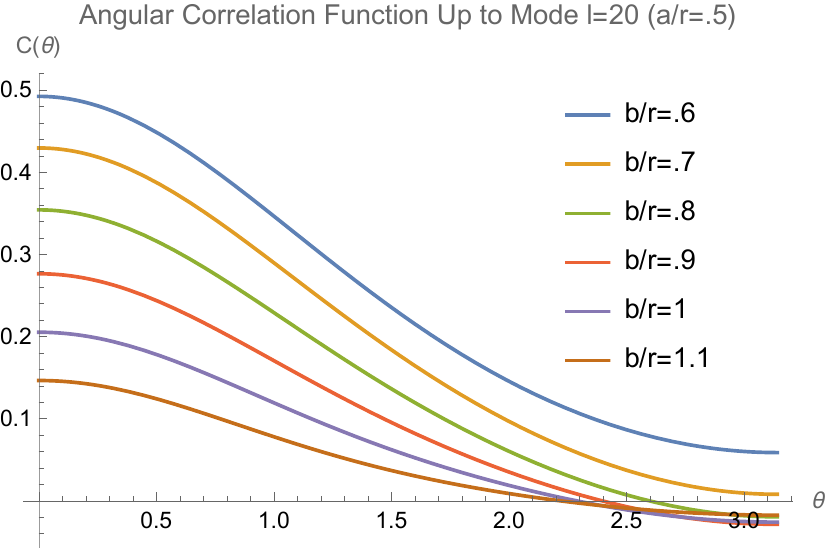}
    \caption{Angular correlation function $C(\Theta)$ for $b\sim R$ and delocalized photon $(a=.5R)$.  }
    \label{fig:midblargeaangcorr}
\end{figure}

For non-zero but small ($b<R$) impact parameter, we must resort to numerical integration. For most ranges of parameters, the dipole moments and quadrupole moments dominate the spectrum. However, for a highly localized photon whose impact parameter is roughly the size of the radius of the sphere of clocks, the spectrum picks up features on smaller angular scales, indicative of higher multipole moments contributing (see \cref{fig:smallbsmallaangcorr}-\cref{fig:midblargeaangcorr}). 

\subsubsection*{ii. Large Impact Parameter}
\begin{figure}[H]
    \centering
    \includegraphics[width=.7\linewidth]{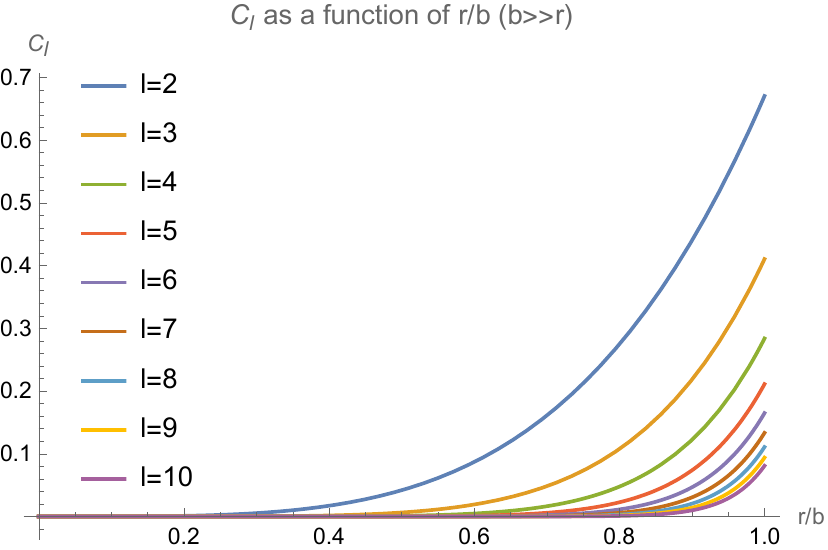}
    \caption{Angular power spectrum $C_{\ell}$ as a function of $r/b$. As impact parameter $b$ increases, higher multipole moments contribute less. }
    \label{fig:largebangpower}
\end{figure}

\begin{figure}[H]
    \centering
    \includegraphics[width=.7\linewidth]{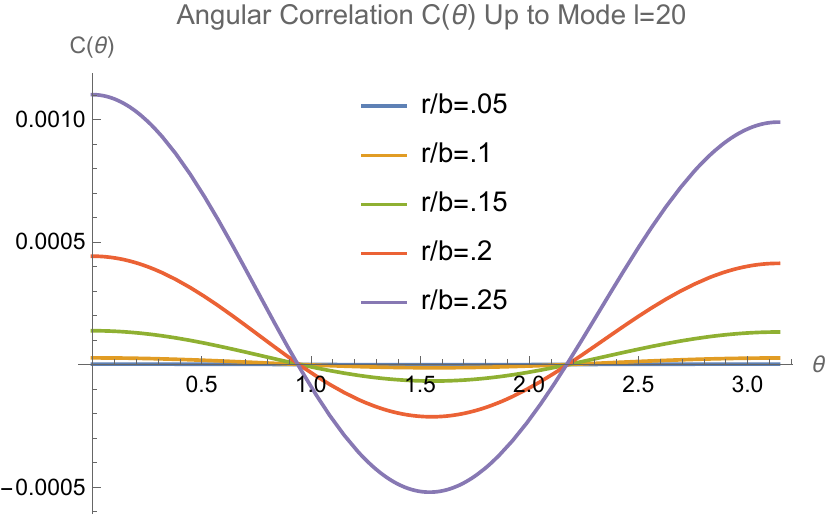}
    \caption{Angular correlation function $C(\Theta)$ summing up to mode $\ell=20$ for various impact parameters. Higher multipole moments only relevant for $b\sim R$. Note that overall amplitude decreases with increasing impact parameter $b$. }
    \label{fig:largebangcorrelation1}
\end{figure}

\begin{figure}[H]
    \centering
    \includegraphics[width=.7\linewidth]{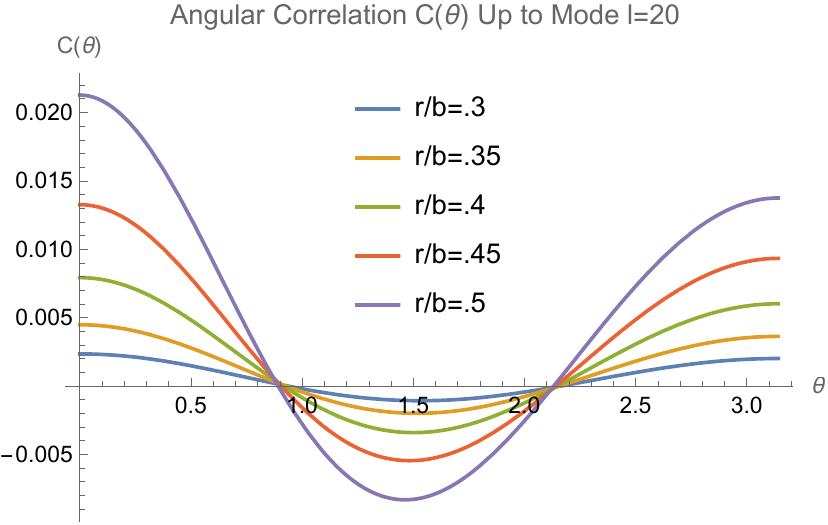}
    \caption{Angular correlation function $C(\Theta)$ summing up to mode $\ell=20$ for various impact parameters. Higher multipole moments are only significant for $b\sim R$. Note that overall amplitude decreases with increasing impact parameter $b$.}
    \label{fig:largebangcorrelation2}
\end{figure}

For the case of large impact parameter, we already have an exact expression for the coefficients $a_{\ell \,m}$ for $b>>R$ in \cref{largebkickseries}. Up to an overall factor of $-4E/b$, we have
\begin{equation}
    a_{\ell \,m} = (-1)^{\ell+1}2^{\ell-1}\ell!\sqrt{\frac{4\pi}{(2\ell+1)!}}((-1)^{\ell}\delta^m_{\ell}+\delta^{m}_{-\ell})\left(\frac{R}{b}\right)^{\ell-1} \label{largebharmoniccoeff}
\end{equation}
Then we get
\begin{equation}
    C_{\ell} = \frac{2^{2\ell+1}}{2\ell+1}\frac{\pi(\ell!)^2}{(2\ell+1)!}\left(\frac{R}{b}\right)^{2(\ell-1)} \label{largebangpower}
\end{equation}
The angular power spectrum $C_{\ell}$ and angular correlation function $C(\Theta)$ are plotted below in \cref{fig:largebangpower}-\cref{fig:largebangcorrelation2}. In general, the quadrupole moment dominates in all cases for large impact parameter (assuming the photon is not so delocalized that its wave function intersects the sphere of clocks). 

\newpage
\subsection{Gravitational Redshift Fluctuation Variance}
Now that we have characterized the angular spectrum of gravitational redshift, we can compute the variance in gravitational redshift fluctuations. Let $\Delta z=\Delta u^r$. We seek formulas of the form
\begin{equation}\label{zvariance}
    \delta_z = \sqrt{\langle(\Delta z)^2\rangle - \langle\Delta z \rangle^2} 
\end{equation}
where $\langle...\rangle$ represents averaging over the sphere. 
\begin{equation}
    \langle f(\theta,\phi)\rangle \equiv 
    \frac{1}{4\pi}\int d\Omega f(\theta,\phi)
\end{equation}
For $b=0$, \cref{0bseriesexpansion} applies. When taking the average over the sphere, only the $\ell=0$ term survives. The result is
\begin{align}
    \langle \Delta z\rangle &= -\frac{4E}{R}\sum_{n=1}^{\infty} \sum_{k=0}^n\frac{(-1)^{n+k+1}}{2^{n}}\frac{1}{k!(n-k)!(2k+1)}\left(\frac{R}{a}\right)^{2n} \nonumber\\
    &=  -\frac{4E}{R^2}(R-\sqrt{2}a D_+(R/\sqrt{2}a))
\end{align}
where $D_+(x)$ is the Dawson integral defined by
\begin{equation}
    D_+(x) = e^{-x^2}\int_0^xe^{t^2}dt
\end{equation}
For large argument, $D_+(x)\sim 1/2x$. Since the spherical harmonics form an orthonormal basis on the sphere, any cross terms when computing $\langle(\Delta z)^2\rangle$ vanish. The result in terms of Dawson integrals is
\begin{equation}
    \langle (\Delta z)^2\rangle = \left(\frac{4E}{R}\right)^2\frac{R + a D_+(R/a) - 2 \sqrt{2} a D_+(R/\sqrt{2} a)}{R}
\end{equation}
Then the fluctuations must scale like
\begin{equation}
    \delta_z^2 = \left(\frac{4E}{R}\right)^2\frac{a (R D_+(R/a) - 2 a D_+(R/\sqrt{2} a)^2)}{R^2} \label{redshiftvar}
\end{equation}
\begin{figure}[H]
    \centering
    \includegraphics[width=.8\linewidth]{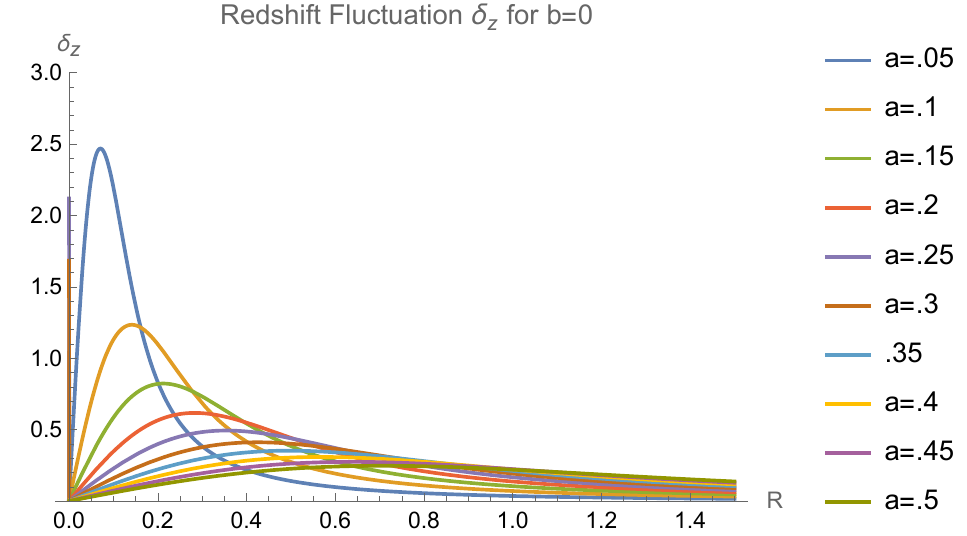}
    \caption{Magnitude of gravitational redshift fluctuations (\cref{redshiftvar}) for zero impact parameter (units normalized by $4E$). Small values of $a$ demonstrate large fluctuations near the photon, while large values of $a$ demonstrate weak fluctuations. The long distance behavior of the fluctuations scale like $Ea/R^2$. }
    \label{fig:0bfluctuations}
\end{figure}

\begin{figure}[H]
    \centering
    \includegraphics[width=.8\linewidth]{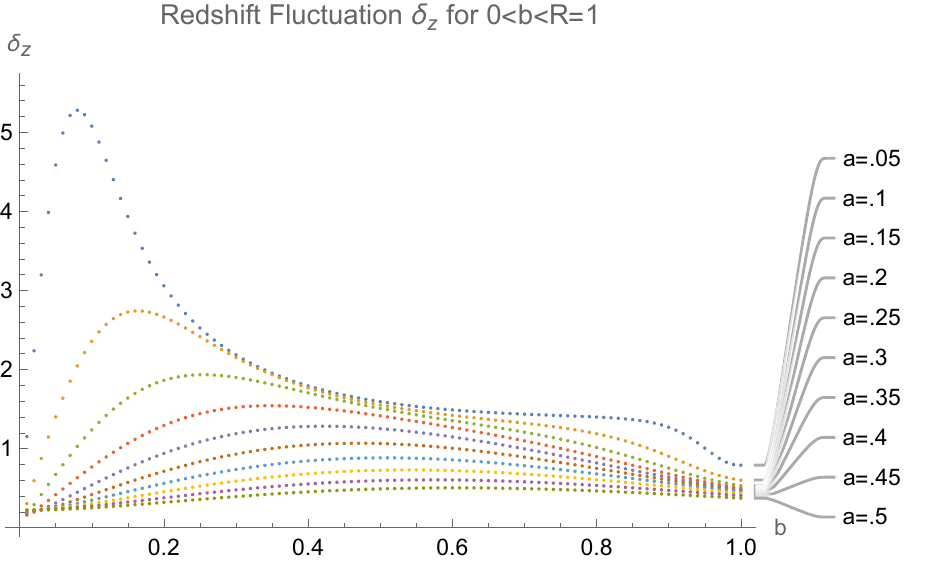}
    \caption{Series of plots for the redshift fluctuation as a function of impact parameter $b$ and localization parameter $a$ for $R=1$. Units are normalized by a factor of $4E/R$.}
    \label{fig:smallbfluc}
\end{figure}

For small impact parameter, numerical integration is needed. The variance in redshift is shown a function of impact parameter for several values of photon localization in \cref{fig:smallbfluc}

For large impact parameters ($b>R$), \cref{largebkickseries} applies. It is immediately apparent that
\begin{equation}
    \langle \Delta z\rangle = 0
\end{equation}
for all values of $\ell$. Since ${\rm cos}(\ell \phi)$ forms an orthogonal function space for $\ell \in \mathbb{Z}$, all cross terms vanish upon integration when computing $\langle (\Delta z)^2\rangle$. After subtraction of the dipole term, we find
\begin{equation}\label{largebvarianceseries}
    \langle (\Delta z)^2\rangle = \left(\frac{4E}{b}\right)^2\sum_{\ell=2}^{\infty}\frac{\sqrt{\pi}\Gamma[\ell+1/2]}{4\Gamma[1+\ell]} \left(\frac{R}{b}\right)^{2(\ell-1)} 
\end{equation}
where $\Gamma[x]=(x-1)!$ is the usual Gamma function. The dominant contribution for $b>R$ is the quadrupole term $\ell=2$. We then find in the $b>>R$ limit
\begin{equation} \label{asympfluc}
     \delta_z \approx \sqrt{\frac{3\pi}{2}}\frac{ER}{b^2}
\end{equation}
If one were to naively sum up contributions from photons over all impact parameters, the result would be divergent (similar to Olber's paradox). The remedy to this issue comes from the fact that distant photons are redshifted, and the gravitational effect of the shockwave they produce scales with the (redshifted) energy of the photon. 

\begin{figure}[H]
    \centering
    \includegraphics[width=.8\linewidth]{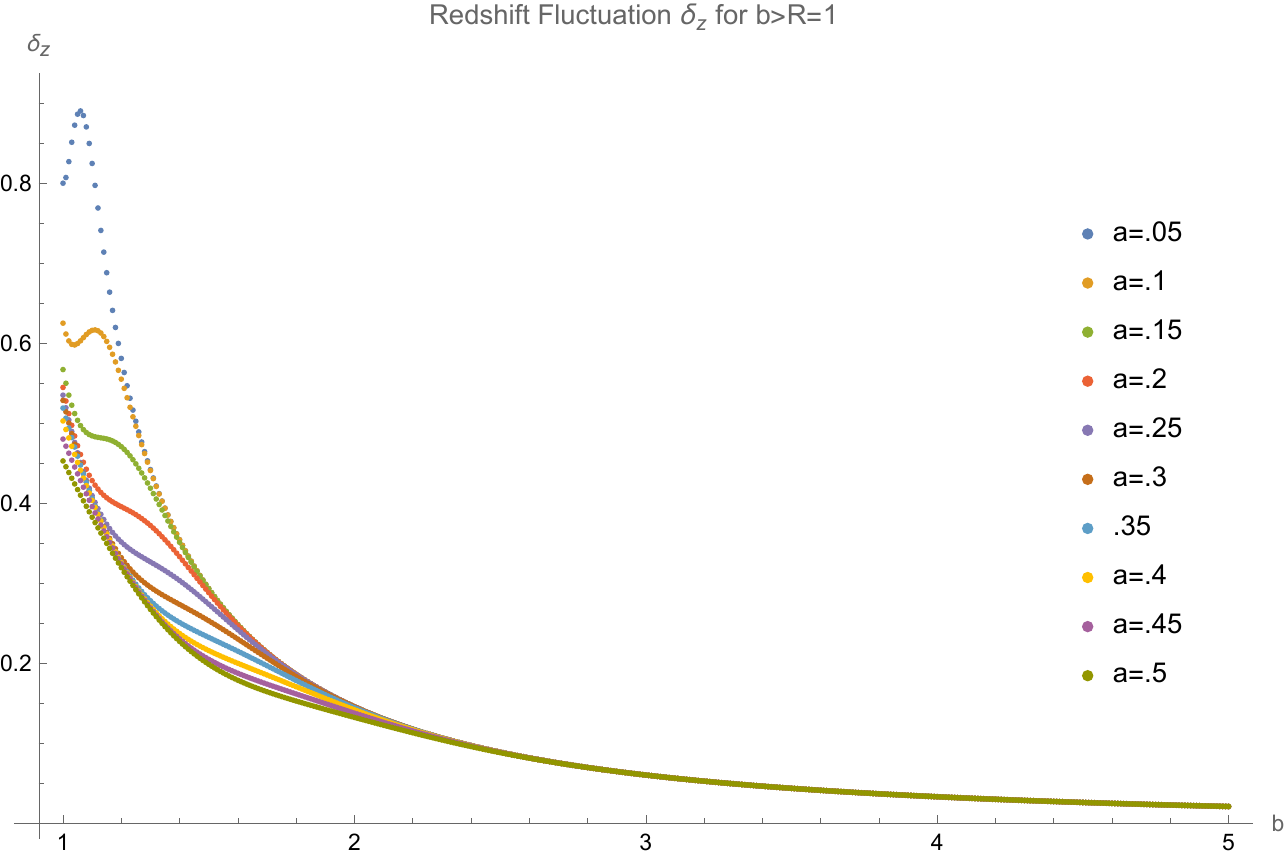}
    \caption{Series of plots of redshift variance (\cref{largebvarianceseries}) as a function of impact parameter $b$ and localization parameter $a$ for $b>R=1$. Units are normalized by a factor of $4E/R$. In the large $b$ limit we see that the fluctuations do not depend on the extent of localization of the photon. The asymptotic behavior is shown in \cref{fig:largebasympfluc}.}
    \label{fig:largebfluc}
\end{figure}
\begin{figure}[H]
    \centering
    \includegraphics[width=.8\linewidth]{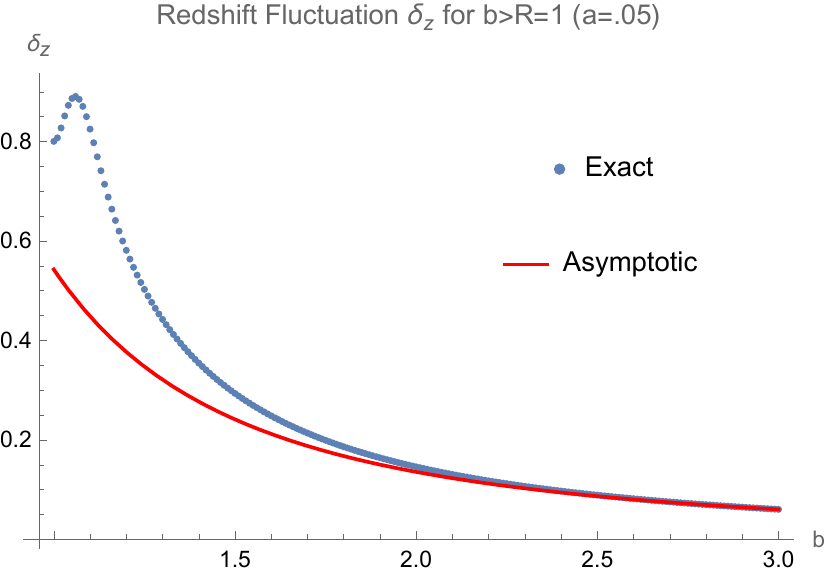}
    \caption{Redshift fluctuations for impact parameter $b>R=1$ for $a=.05$. Blue data points show true numerical data, while the red curve shows the asymptotic behavior consistent with \cref{asympfluc}. Units normalized by a factor of $4E/R$. }
    \label{fig:largebasympfluc}
\end{figure}

\subsection{Time Dependence and Frequency Spectrum of Redshift Anisotropy}
Finally, we seek to characterize the time evolution of the redshift of a clock at a fixed location on the sphere as measured by the central observer. To do this, we need to compute the retarded time associated with the intersection of the planar shockwave at a given point on the sphere.
\begin{figure}[t]
    \centering
    \includegraphics[width=.6\linewidth]{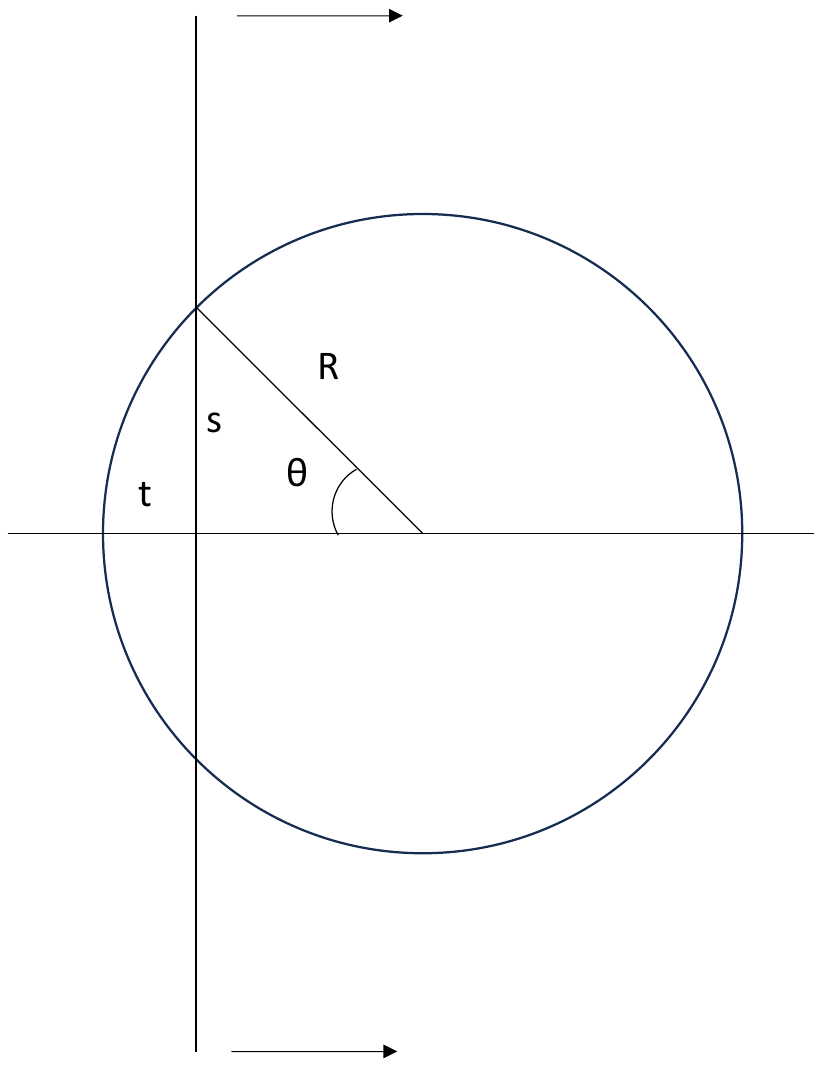}
    \caption{Visual representation of a planar shockwave crossing the sphere. The time of intersection can be defined as a function of the angle $\theta$. The geometry is the same for photons with $0$ or finite impact parameter since the gravitational shockwave is planar in both cases.}
    \label{fig:shockcrossing}
\end{figure}

Define the intersection time of the plane with a point on the sphere at angle $\theta$ as $t_{\rm int}$. We will choose our time coordinate such that $t_{\rm int}=0$ when the shockwave first touches the sphere at its pole. Next, define the time at which the observer at the center of the sphere observes the intersection occurring to be $t_{\rm obs}$. Then we have
\begin{equation}
    t_{\rm int} = 2 R {\rm sin}^2(\theta/2) 
\end{equation}
\begin{equation}
    t_{\rm obs}= t_{\rm int}+ R = R(2-{\rm cos}\theta)
\end{equation}
\begin{figure}[H]
    \centering
    \includegraphics[width=\linewidth]{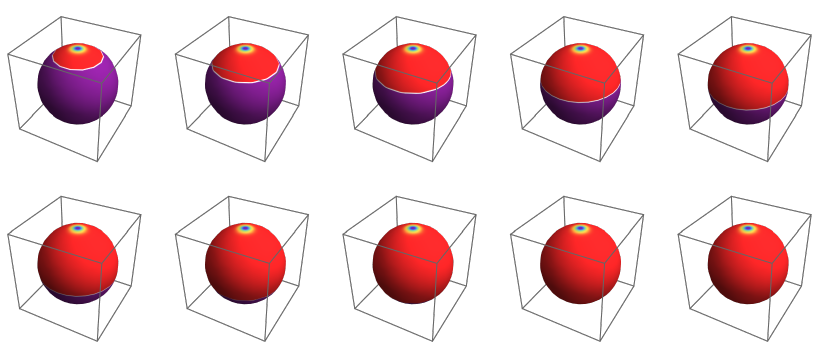}
    \caption{Time evolution of heat map representing gravitational redshift for zero impact parameter. Time increases moving from left to right, top to bottom. }
    \label{fig:0btimeevolution}
\end{figure}
\begin{figure}[H]
    \centering
    \includegraphics[width=\linewidth]{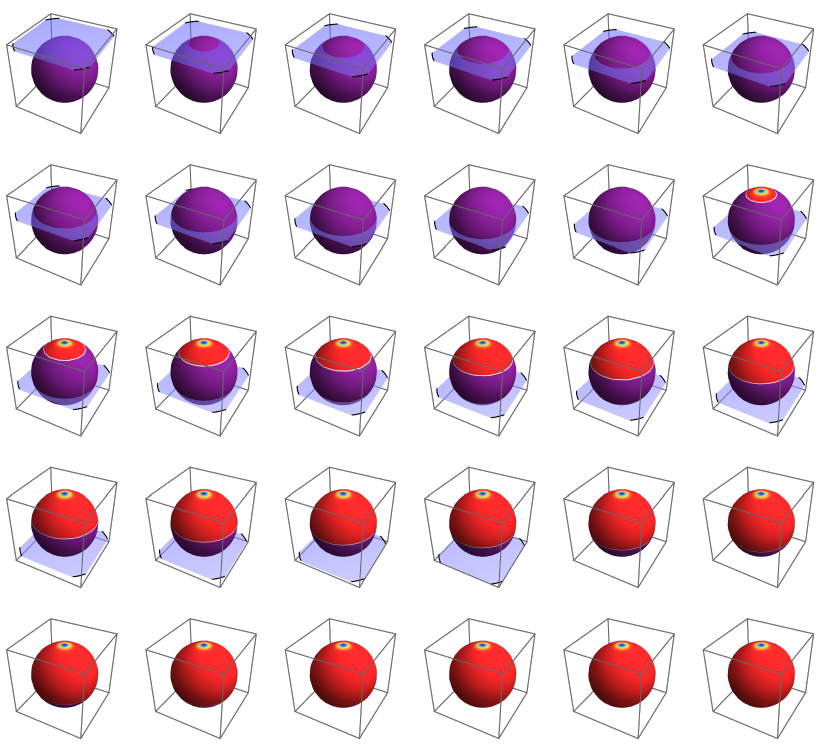}
    \caption{Time evolution of gravitational shock wave and the resulting heat map representing the gravitational redshift. Due to causal constraints, there is a delay between the time of the shock reaching a given clock on the sphere and the time the central observer is able to see the resulting redshift.}
    \label{fig:shockwaveheatmaptimeevolution}
\end{figure}
\begin{figure}[H]
    \centering
    \includegraphics[width=\linewidth]{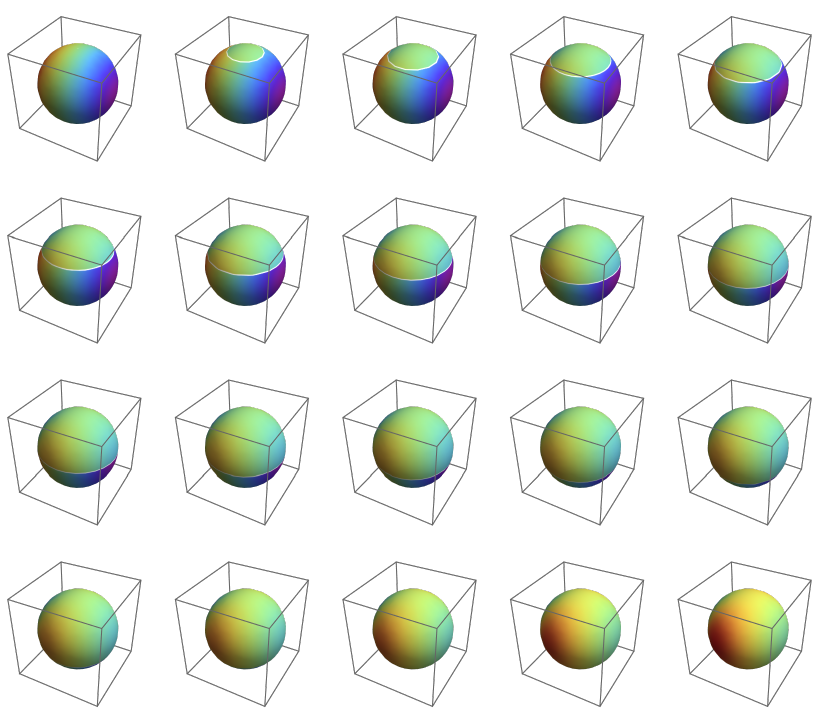}
    \caption{Time evolution of heat map representing gravitational redshift for impact parameter $b>R=1$. Between $t=1$ and $t=2$, the angular spectrum is purely dipolar for $\theta> {\rm cos}^{-1}(2-t)$ and predominantly quadrupolar for $\theta< {\rm cos}^{-1}(2-t)$.}
    \label{fig:smallbheatmaptimeevolution}
\end{figure}

Now we would like to understand how the measurable effects on the sphere vary with time and derive a temporal spectrum. Regardless of the relative size of the impact parameter and the relative localization of the photon, the shock wave will always cross the sphere in a uniform plane. When the shock wave initially hits the sphere at the pole, the central observer has not yet had time to see that the clocks have been redshifted at this point. It is only the moment that the shockwave crosses the observer that they can first begin to see the clocks' redshift. As the shockwave sweeps over the sphere, the intersection of the shockwave with the sphere forms a cone whose deficit angle changes with time (not to be confused with the usual null light cone). The time at which the observer sees a clock redshift is the retarded time associated with that event. Clocks that are outside of this cone after the moment the observer has experienced a velocity kick will effectively be seen by the observer to have redshifted due to the motion of the observer, and the angular spectrum of this shift is purely dipolar for zero impact parameter. Once enough time has elapsed, this dipole term will be eliminated, and the quadrupole and higher order multipoles will remain. In general, the Fourier transform for a step function is given by
\begin{align}
    F[\theta(t-t_0)] & = \frac{1}{2\pi}\int_{-\infty}^{\infty} \theta(t-t_0)e^{-i\omega t}dt \\
    &= \frac{i e^{i t_0 \omega}}{2\pi \omega}+\frac{1}{2}\delta(\omega)
\end{align}
Similarly, if we have a finite square pulse, we get
\begin{equation}
    F[\theta(t-t_0)\theta(t_1-t)] = \frac{i(e^{it_0\omega}-e^{it_1\omega})}{2\pi\omega}
\end{equation}
In general, sudden changes in the temporal domain correspond to $1/\omega$ amplitudes in the frequency domain. Even if the change is spread out over a finite time, the low frequency limit is dominated by the $1/\omega$ term. For example, if we consider a system whose rate of change in the temporal domain is given by a Gaussian, then the accumulate change is given by
\begin{equation}
    \Delta = \int_{-\infty}^t \frac{e^{-x^2/(2\tau^2)}}{\sqrt{2\pi\tau^2}}dx = \frac{1}{2}\left(1+{\rm Erf}\left(\frac{t}{\sqrt{2}\tau}\right)\right)
\end{equation}
where ${\rm Erf}(x)$ is the exponential error function integral. In the frequency domain, this looks like
\begin{equation}
    F[\Delta] = \frac{i e^{-\tau^2 \omega^2/2}}{2\pi \omega}+\frac{1}{2}\delta(\omega)
\end{equation}

\section{Anisotropic Time Shift}

In addition to gravitational redshift, the passage of a shock generates a displacement or time shift. It does not involve any exchange of energy, but still produces observable effects during the passage of a shock.

The instantaneous relative displacement shift due to a spherically symmetric shockwave produced by the decay of a massive particle is given by \cite{Mackewicz_2022}.
\begin{equation}
    \Delta D^a = \frac{M}{r}(\theta_a\theta_b-\phi_a\phi_b)D^b = \frac{\Delta_{ab}}{r}D^b
\end{equation}
This relative displacement kick is associated with instantaneous time translations on a system of spherically arranged clocks.
\begin{equation}
    \Delta_{ab}= \left(\mathcal{D}_a\mathcal{D}_b-\frac{1}{2}q_{ab}\mathcal{D}^2\right)T(\theta,\phi)
\end{equation}
$T(\theta,\phi)$ represents a shift in the retarded time coordinate of the clock, and $\mathcal{D}_a$ represents covariant angular derivatives on the sphere. 

One can easily verify that the following shift as a function of zenith angle $\theta$ gives the correct memory tensor $\Delta_{ab}$.
\begin{equation}
    T(\theta)=E\left((1-{\rm cos}\theta){\rm ln}(1-{\rm cos}\theta)+(1+{\rm cos}\theta){\rm ln}(1+{\rm cos}\theta)+1-2{\rm ln}2\right) \label{sphereshift}
\end{equation}
Decomposition into Legendre polynomials gives an expression consistent with the result of \cite{Mackewicz_2022}, namely that the spectrum is predominantly quadrupolar in nature, with some small corrections near the poles $\theta=0,\pi$ due to the higher $\ell$ modes. 

We are interested in using this result in the case of a planar (or nearly planar) shockwave. Mathematically speaking, this is the limit $d>>\rho = d {\rm sin}\theta$, or $\theta<<1$,where $d$ is the distance the photon has traveled from its creation. Taylor expanding \cref{sphereshift} around $\theta=0$ gives
\begin{equation}
    T(0+\delta \theta)=E\left(1-\left(\frac{1}{2}-{\rm ln}(\delta \theta^2/4)\right)\delta \theta^2+\mathcal{O}(\delta\theta^4)\right)
\end{equation}
We can therefore see that in the strict planar limit, the time shift is constant. This is in disagreement with the result of \cite{DRAY1985173}, but is consistent with the fact that the curvature of an infinite planar shock wave has no derivative of delta function term, which is necessary for memory and a relative displacement kick/time shift. If we consider a slightly curved, nearly planar shockwave, the time shift in terms of the transverse distance from the photon $\rho$ looks like
\begin{equation}
    T(\rho)=E\left(1-\frac{1}{2}\frac{\rho^2}{d^2}\right)
\end{equation}

Since the time shift is constant for all observers in the limit of a planar shock, the shift can be removed by a gauge transformation and an observer will not be able to measure a relative difference in their clock compared to the system of spherically arranged clocks after the photon as passed the entire system. However, due to causal constraints, there will be a window of time of the size of the light crossing time when the observer \textit{can} measure a relative difference between clocks, due to the fact that some of the clocks have not yet registered a shift according to the observer. In the time domain, this behavior is a rectangular pulse. We are interested in the frequency power spectrum of the gravitational effects measured by the observer. Consider a function $f(t)$. We can define its convolution as
\begin{equation}
    \bar{f}(\tau) = \int_{-\infty}^{\infty} f(t-\tau)f(t)dt
\end{equation}
The frequency power spectrum is then given by
\begin{equation}
    P(\omega) = \int^{\infty}_{-\infty} \bar{f}(\tau)e^{-i\omega \tau}d\tau
\end{equation}
For a rectangular pulse, the convolution is a triangular pulse, whose Fourier transform is the square of a sync function (sin$x$/$x$). 

The width of the rectangular pulse for a fixed angle as measured by the central observer is determined by the interval between the time the shockwave hits the observer and one radius travel time $R$ \textit{after} the shockwave hits a clock sitting at the angle of interest. The width for the head-on clock ($\theta=0$) is therefore zero, and the width for the trailing clock is $2R$. 
\begin{equation}
    \delta t (\theta) = R(1-{\rm cos}\theta)
\end{equation}
The frequency power spectrum for a given angle $\theta$ is then given by
\begin{equation}
    P(\omega,\theta) = \frac{E^2}{ \sqrt{2\pi}}{\rm sinc}^2\left(\frac{\omega R}{2}(1-{\rm cos}\theta)\right)
\end{equation}

\section{Pure phase anisotropy and vacuum fluctuations}

As noted above, displacements of relative clock  position carry no energy.  However, they would be measured as fluctuations of differential phase in an interferometer with mirrors at the locations of the clocks.
As with the fluctuations from a real photon gas, the angular and temporal spectra of virtual fluctuations should depend on the space-time structure of the measurement, with a normalization determined by an energy cutoff of the vacuum fluctuation state.
We can use  zero-energy differential gravitational phase displacements from photon shocks  as an indication of   causal constraints on  temporal and angular  phase distortions from  fluctuations associated with null vacuum states.

\begin{figure}[H]
    \centering
    \includegraphics[width=0.8\linewidth]{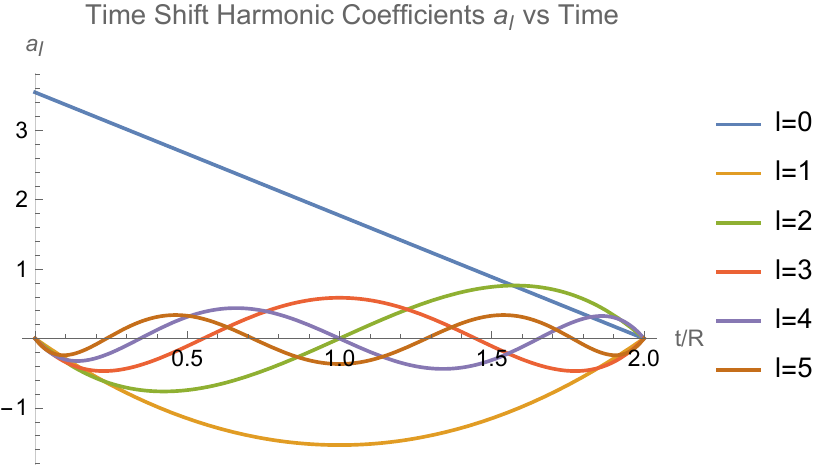}
    \caption{Time evolution of angular harmonic components of zero-energy clock phase displacement during a shock passage (units normalized by factor of $E$). Angular profile is a discontinuous jump at $\theta = {\rm cos}^{-1}(1-t/R)$ (\cref{timeshiftanisotropy}). A basis is chosen along the propagation, so the only nonzero harmonics have $m=0$ for all $\ell$. }
    \label{harmonicsintime}
\end{figure}

During a  shock passage, clocks on one side of the shock are uniformly displaced from those on the other.  The evolution of angular harmonic components during this passage is shown in fig. \ref{harmonicsintime}.  The monopole linearly decreases from its initial to final value, representing the total net displacement from the shock. The dipole component reaches a maximum value at the halfway point.   Other low-order harmonics vary more rapidly with time, with amplitudes that fall off with wavenumber. 
\begin{equation}\label{timeshiftanisotropy}
    T(\theta,t) = E \theta(1-{\rm cos}\theta-t/R)
\end{equation}

\begin{figure}[H]
    \centering
    \includegraphics[width=0.8\linewidth]{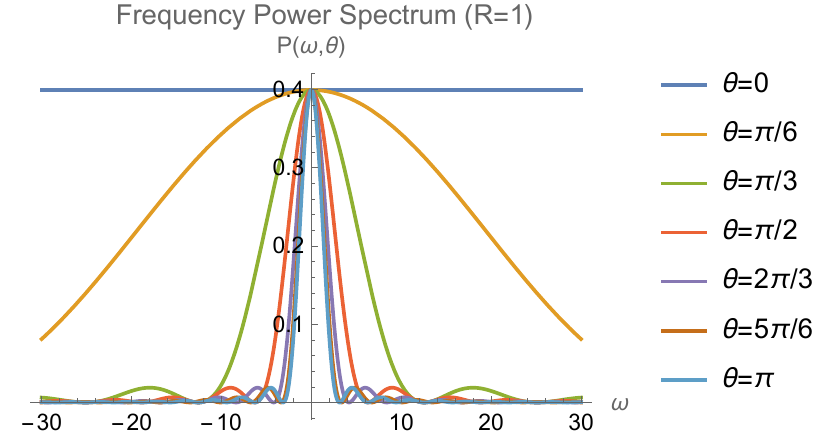}
    \caption{Frequency power spectrum for relative temporal shift between a clock at angle $\theta$ on the sphere and the central observer (units normalized by $E^2$). Size of sphere is set to 1 for visualization. Effective width of power spectrum is inversely proportional to light crossing time of the sphere of clocks.  }
    \label{frequencypower}
\end{figure}

The corresponding temporal frequency spectra are shown in fig. \ref{frequencypower}. They show  spectra characteristic of those that would appear in the signal of an interferometer, with a weighting that depends on the angular configuration of the mirrors. The time scale is set by the length scale of the mirror spacing. A superposition of many virtual shocks, such as a  vacuum state up to some energy scale prepared at infinity,  would produce the same power spectra in time and angle, but with larger amplitude.

\begin{figure}[H]
    \centering
    \includegraphics[width=\linewidth]{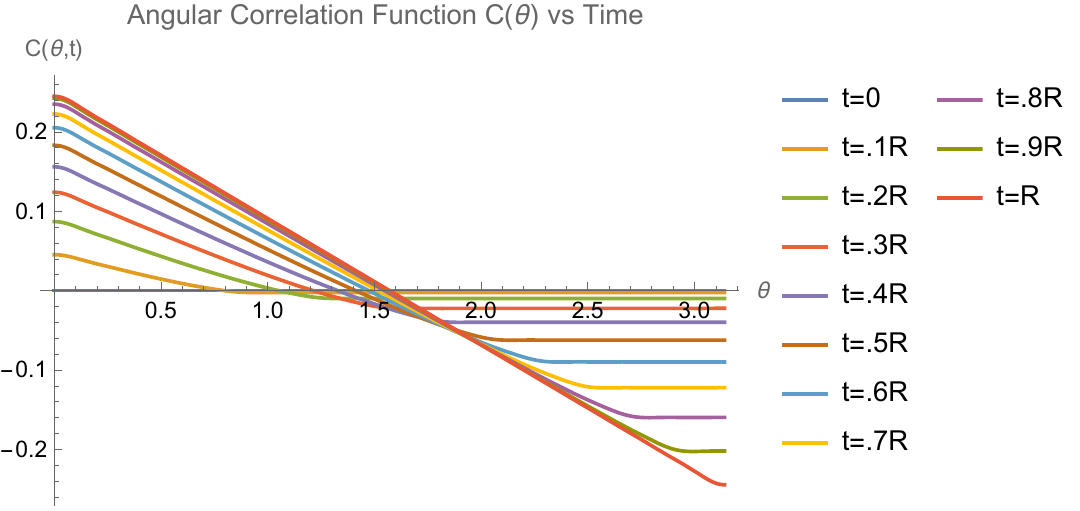}
    \caption{Angular correlation of phase displacement during the first half of a shock passage. The second half is the same in reverse. The sharp angular boundary and linear behavior of the correlation result from constraints of causal symmetry, which should  apply to vacuum fluctuations.}
    \label{angularcorrelations}
\end{figure}

The  symmetry of the  shock is manifest in the 
 angular correlation function $C(\Theta)$ as the shock advances, shown in fig. \ref{angularcorrelations}.  Because of the planar symmetry of a shock prepared at infinity, harmonics of all orders ``conspire''  to produce a linear angular correlation. As expected, the maximum correlation occurs at the center of the shock passage, when it has a purely odd parity. The same shape for $C(\Theta)$ is produced by  the pure-phase component of noise from a gas of real photons prepared at distances much larger than the size of the measurement apparatus. Similar causal constraints should apply to fluctuations from vacuum states.

If we take these results as estimates of the phase noise produced by virtual null particles with a cutoff at the Planck  scale, they roughly
 accord with previous estimates from other methods of the timescale and magnitude of interferometric phase fluctuations produced by causally-coherent quantum gravitational vacuum fluctuations\cite{Hogan2007,Hogan2008,Hogan2012,Kwon:2014yea,Kwon2022,VERLINDE2021,BanksZurek2021,Verlinde2022,banks2023fluctuations}.
For a UV cutoff at the Planck scale, it  appears likely that they can be   measured with current technology \cite{Holo:Instrument,holoshear,Richardson2021,vermuelen2020,vermeulen2024photon}.
However, the  correlations of fluctuations represented  in these plots highlights the importance of  geometrical  layout of interferometer experiments, because causal symmetry creates ``conspiracies'' of harmonic components, which could  make some 
configurations  intrinsically insensitive to signals. 
For example, note that the  angular correlation for separations $\Theta\sim 90^{\circ}$ is significantly smaller than the overall correlation amplitude.  This behavior differs  from that of the redshift anisotropy discussed above from real photons, which is dominated by the quadrupole.  The corresponding signals in an experiment with a pure right-angle configuration differ by an order of magnitude.
Measurements of angular and temporal spectra and correlation functions with a variety of interferometer layouts could  provide
 detailed maps of the space-time structure and coherence and  of  gravitational quantum fluctuation states.

\section{Conclusion}
It has been shown  that with  appropriate averaging of the metric perturbation, Riemann curvature, and velocity kick experienced by a test body due to the passing of a massless point particle, the mean gravitational effect of many null particles is consistent with that of a photon gas. In other words, we have demonstrated that  the acceleration experienced by a test body due to a homogeneous and isotropic perfect null fluid can be thought of as a sum  of randomly oriented instantaneous velocity kicks due to the passage of gravitational shockwaves. 

We have derived  angular profiles and spectra of gravitational redshifts measured by a central observer on a system of spherically arranged clocks for a range of transverse particle profiles and impact parameters. As one example, we have shown that after correctly accounting for the motion of the central observer, the angular spectrum for large impact parameter ($b>R$) is predominantly quadrupolar in nature. In general, the spectra of fluctuations are determined by the spacetime distributions of the measuring system and particle wave functions, with a normalization determined by the numbers and energies of particles.

We have also separated the transient part of the shock  effect that transfers no energy or momentum, and leaves no permanent imprint behind. These transient angular perturbations in clock displacement or phase were used to illustrate behavior 
that might characterize the response of an idealized macrocopic interferometer  to quantum vacuum fluctuations of gravity, which should  preserve similar causal symmetries of angular correlation. These results  suggest that  experiments should be  capable of exploring a variety of  geometrical configurations in order to detect and characterize quantum fluctuations of the gravitational vacuum.

\bibliography{photongas}

\section{Appendix}

\subsection{Curvature Identities}
In this appendix, we summarize curvature identities that are useful in simplifying some of the calculations performed. For a traceless stress energy tensor, the Einstein equations reduce to
\begin{equation}
    R=-8\pi T = 0 
\end{equation}
\begin{equation}
    R_{ab}= 8\pi T_{ab}
\end{equation}
The Bianchi identity can be written in the following three ways.
\begin{equation}
    \nabla_{[a}R_{bc]de}=0
\end{equation}
\begin{equation}
    \Box R_{bcde} = -\nabla_a \nabla_b R_{cade} - \nabla_a \nabla_c R_{abde} 
\end{equation}
\begin{equation}
    \nabla^aR_{abcd} = \nabla_c R_{db}-\nabla_dR_{cb}
\end{equation}
The relationship between Riemann curvature, Ricci curvature, and Weyl curvature is given by
\begin{equation}
    R_{abcd}=C_{abcd} + \frac{1}{2}(R_{ac}g_{bd}-R_{ad}g_{bc}-R_{bc}g_{ad}+R_{bd}g_{ac})+\frac{R}{6}(g_{ad}g_{bc}-g_{ac}g_{bd})
\end{equation}
The difference between mixed covariant derivatives of a tensor field is related to the curvature by
\begin{equation}
    (\nabla_a\nabla_b-\nabla_b\nabla_a) T^{c_1...c_m}_{d_1...d_n} = -\sum_i R_{abe}^{\quad c_i} T^{c_1...e...c_m}_{d_1...d_n} + \sum_j R_{abd_j}^{\quad e} T^{c_1...c_m}_{d_1...e...d_n}
\end{equation}
In the linearized theory, we have that
\begin{equation}
    (\nabla_a\nabla_b-\nabla_b\nabla_a)R_{cdef} = 0
\end{equation}
\begin{equation}
    \Box R_{abcd} = 32\pi \nabla_{[a}\nabla_{|[d}T_{c]|b]}
\end{equation}

\end{document}